\documentclass[english,aps,prc,nofootinbib,superscriptaddress,showpacs,twocolumn,letter]{revtex4}
\usepackage{amsmath}
\usepackage[    bookmarks,
                 bookmarksopen = true,
                 bookmarksnumbered = true,
                 linktocpage,
                 colorlinks = true,
                 linkcolor = blue,
                 urlcolor  = blue,
                 citecolor = blue,
                 anchorcolor = green,
                 hyperindex = true,
                 hyperfigures]
		{hyperref}
\usepackage[utf8]{inputenc}
\usepackage{graphicx}
\usepackage{esint}
\usepackage{xcolor}

\makeatletter
\@ifundefined{textcolor}{}
{%
 \definecolor{BLACK}{gray}{0}
 \definecolor{WHITE}{gray}{1}
 \definecolor{RED}{rgb}{1,0,0}
 \definecolor{GREEN}{rgb}{0,1,0}
 \definecolor{BLUE}{rgb}{0,0,1}
 \definecolor{CYAN}{cmyk}{1,0,0,0}
 \definecolor{MAGENTA}{cmyk}{0,1,0,0}
 \definecolor{YELLOW}{cmyk}{0,0,1,0}
}
\makeatother

\usepackage{babel}

\begin{document}

\preprint{This line only printed with preprint option}

\title{
Magnetic field-induced squeezing effect at RHIC and at the LHC
}

\author{Long-Gang Pang}
\affiliation{Frankfurt Institute for Advanced Studies, Ruth-Moufang-Strasse 1, 60438 Frankfurt am Main, Germany}
\author{Gergely Endr\H{o}di}
\affiliation{Institute for Theoretical Physics, University Regensburg, D-93040 Regensburg, Germany}
\affiliation{Institute for Theoretical Physics, Goethe University, Max-von-Laue-Strasse 1, 60438 Frankfurt am Main, Germany}
\author{Hannah Petersen}
\affiliation{Frankfurt Institute for Advanced Studies, Ruth-Moufang-Strasse 1, 60438 Frankfurt am Main, Germany}
\affiliation{Institute for Theoretical Physics, Goethe University, Max-von-Laue-Strasse 1, 60438 Frankfurt am Main, Germany}
\affiliation{GSI Helmholtzzentrum f\"ur Schwerionenforschung, Planckstr. 1, 64291 Darmstadt, Germany}

\selectlanguage{english}%

\begin{abstract}
In off-central heavy-ion collisions, the quark-gluon plasma (QGP) 
is exposed to the strongest magnetic fields ever created in the universe.
Due to the paramagnetic nature of the QGP at high temperatures, 
the spatially inhomogeneous magnetic field configuration exerts 
an anisotropic force density that competes with the pressure gradients 
resulting from purely geometric effects. 
In this paper, we simulate (3+1)-dimensional ideal
hydrodynamics with external magnetic fields to estimate the effect
of this force density on the anisotropic expansion of the QGP in
collisions at RHIC and at the LHC.
While negligible for quickly decaying magnetic fields, we find 
that long-lived fields generate a substantial force density that 
suppresses the momentum anisotropy of the plasma by up to $20\%$ at 
the LHC energy, and also leaves its imprint on the elliptic flow $v_2$ of
charged pions. 
\end{abstract}
	
\keywords{Relativistic Heavy-ion collisions, magnetic field, elliptic flow}
\pacs{12.38.Mh,25.75.Ld,25.75.Gz}

\maketitle

\section{Introduction}

One of the most striking phenomena in high-energy heavy-ion collisions
is the strong collective flow of the hot quark-gluon plasma (QGP), caused  
by the relativistic hydrodynamic expansion driven by local pressure gradients. 
For off-central collisions, the anisotropy of the initial geometry 
enhances this collective flow in the reaction plane
and leads to a non-vanishing elliptic flow $v_2$ of charged hadrons
observed in the experiment.
An additional relevant feature of such collisions is the generation of 
extremely strong magnetic fields~\cite{Skokov:2009qp,Voronyuk:2011jd,Bzdak:2011yy,Deng:2012pc,Kharzeev:2012ph}
created by the colliding charged beams moving at relativistic speed. 
The impact of these fields in heavy-ion collisions 
was first explored in relation with the 
chiral magnetic effect~\cite{Kharzeev:2007jp,Fukushima:2008xe}.
Later studies also discussed the role of magnetic fields
on jet energy loss~\cite{Tuchin:2010vs}, on the thermal photon and di-lepton production 
rate~\cite{Tuchin:2012mf,Tuchin:2013ie}
and on the collective expansion of the 
fireball~\cite{Mohapatra:2011ku,Bali:2013owa}.

In this paper we concentrate on the response of the QGP to the magnetic field 
in terms of the induced magnetization,
\begin{equation}
  \mathbf{M} \equiv -\frac{\partial f}{\partial
  \mathbf{B}
  },
  \label{eq:magnetization}
\end{equation}
where $f$ is the free energy density of the thermodynamic system.
Lattice QCD calculations have demonstrated that 
in the range of 
magnetic fields and temperatures relevant for heavy-ion collisions
$\mathbf{M}\parallel \mathbf{B}$, that is to say, the QGP behaves as a 
paramagnet\footnote{
Previously, the paramagnetic response was also predicted in 
perturbation theory~\cite{Elmfors:1993wj} and in the 
hadron resonance gas model~\cite{Endrodi:2013cs}.}~\cite{Bali:2013esa,Bali:2013txa,Bonati:2013lca,Bonati:2013vba,Levkova:2013qda,Bali:2014kia}. 
It is well known that, if exposed to inhomogeneous magnetic fields, 
paramagnetic materials move along the gradient of $|\mathbf{B}|$ -- as opposed to diamagnets
that move in the opposite direction~\cite{landau:1995em}. 
Based on model descriptions of heavy-ion collisions (see below), 
off-central events are expected to exhibit inhomogeneous fields with a strong spatial anisotropy.
As Ref.~\cite{Bali:2013owa} pointed out, such a magnetic field configuration 
exerts an anisotropic force density on the QGP that compresses 
matter in the transverse plane. This force density -- dubbed {\it 
paramagnetic squeezing} --
arises as the system minimizes its free energy and reads
\begin{equation}
  \mathbf{F} \equiv -\nabla f = (\nabla \mathbf{B}) \cdot \mathbf{M}. 
  \label{eq:force_density}
\end{equation}
It was recognized that the squeezing force density might affect 
the collective expansion and, eventually, 
have an impact on the elliptic flow $v_2$
of the final charged hadrons.
A first estimate of this effect was made 
by comparing the magnitudes of the squeezing force density
with those of the pressure gradients at initial time for RHIC and LHC energies \cite{Bali:2013owa},
revealing that the effect might be 
marginal for RHIC but substantial for LHC collisions.

In this paper we improve on this simplistic estimate in various aspects. 
We simulate (3+1)-dimensional ideal hydrodynamics to
determine the time-evolution of the energy density and the fluid velocity in the 
presence of an external magnetic field profile.
This profile is assumed to be given in terms of a few parameters 
that control the magnitude, the spatial distribution and the 
time-evolution of the magnetic field.
Systematically varying these parameters enables us to study the influence of 
paramagnetic squeezing in different limits.
The effect of the squeezing force density is taken into account
throughout the hydrodynamic expansion to determine the impact on
momentum anisotropy and to calculate the time integrated effect on $v_2$ of charged particles on the freeze out 
hypersurface. 

Note that in this setup we treat the magnetic field as an external degree 
of freedom, i.e., we neglect the back-reaction of the fluid on $\mathbf{B}$. 
A fully consistent description of the entangled evolution of the hydrodynamic 
expansion and of the electromagnetic field would require (3+1)-dimensional 
relativistic magnetohydrodynamics, which, however, is still under development~\cite{Lyutikov:2011vc,Roy:2015kma}.
We further mention that magnetic field-induced effects on the fluid expansion
have also been discussed in terms of a comparison 
of the magnetic and fluid energy densities~\cite{Mohapatra:2011ku,Roy:2015coa}
and of the 
one-dimensional longitudinal
boost-invariant Bjorken flow~\cite{Pu:2016ayh,Pu:2016bxy}.

The rest of this paper is organized as follows. 
In Sec.~\ref{sec:method} we introduce the (3+1)-dimensional ideal
hydrodynamic model and the parameters used for the magnetic field and the hydrodynamic expansions.
The relative magnitude of the squeezing force density and  
the initial pressure gradients, together with the effect of the squeezing on 
the fluid expansion and on the elliptic flow of final charged hadrons are shown
in Sec.~\ref{sec:results} both for Pb+Pb $\sqrt{s_{NN}}=2.76$ TeV collisions and for 
Au+Au $\sqrt{s_{NN}}=200$ GeV collisions.
This is followed by Sec.~\ref{sec:summary}, which summarizes the results 
and presents a short outlook.

\section{Setup and methods \label{sec:method}}

We use (3+1)-dimensional ideal hydrodynamic 
simulations~\cite{Pang:2012he,Pang:2014ipa}, as implemented in the 
CLVisc code parallelized on GPUs using OpenCL~\cite{Pang:2014ipa}.
For simplicity, no viscous corrections are taken into account
in the current study.

\subsection*{Hydrodynamic equations}

%\cite{Mallick:2011ar}.
The hydrodynamic equations
in the presence of the squeezing force~(\ref{eq:force_density}) read~\cite{Felderhof:1999Mhd}
\begin{equation}
  \partial_{\mu}T^{\mu\nu} = F^{\nu}, \label{eq:magnetic_fluid}
\end{equation}
where $T^{\mu\nu}=(\varepsilon+P)u^{\mu}u^{\nu}-Pg^{\mu\nu}$
is the energy momentum tensor for ideal hydrodynamics, $u^{\mu}=\gamma(1,\mathbf{v})$ denotes the
fluid velocity four vector, $\varepsilon$ the energy density and
$P$ the pressure\footnote{
Notice that the energy-momentum tensor of the magnetic field -- i.e., terms like 
$\mathbf{B}^2/2$ in the energy density -- are not taken into account here. 
This is due to the fact that the field is considered external so that, for example, its 
energy density is independent of the fluid dynamics. The only effect appears due to the 
interaction of $\mathbf{B}$ with the fluid through the magnetization.
}. The latter is given as a function of $\varepsilon$ by the
equation of state (EoS).
The lattice QCD equation of state from the Wuppertal-Budapest group (2014) \cite{Borsanyi:2013bia} is used in the current study.
The space-time coordinate axes are $x$ (impact parameter direction), $y$ 
(direction perpendicular to the reaction plane), $\eta_s$ (space-time rapidity) 
and $\tau$ (proper time).

According to lattice QCD calculations~\cite{Bali:2013owa,Bali:2014kia}, 
the leading expansion for the magnetization -- $\mathbf{M}=\chi\mathbf{B}$ 
in terms of the magnetic susceptibility $\chi$ --
is a reasonable approximation for the range of magnetic fields
and temperatures we are interested in. Using the susceptibility, 
the squeezing force density $F^{\nu}$ on the right hand side 
of Eq.~(\ref{eq:magnetic_fluid}) reads
\begin{equation}
F^{x} = \frac{\chi}{2}\,\partial_{x}|\mathbf{B}|^2, \quad\;
F^{y} = \frac{\chi}{2}\,\partial_{y}|\mathbf{B}|^2 , \quad\;
F^{\tau} = F^{\eta_s} = 0.
\end{equation}
Here we assumed that the dynamics relevant for the anisotropic flow 
is governed by forces in the transverse plane and, accordingly, set the 
longitudinal force to zero. This choice may also be thought of as a setup
where the magnetic field is constant in the longitudinal direction within the QGP.
In addition, we have checked that
enforcing zero energy input $u_\mu F^\mu=0$ from the force (as expected in 
a purely magnetic background) by
setting the zero component $F^{\tau}$ to $v_x F^{x} + v_y F^{y}$ 
changes the momentum anisotropy by only $\sim1\%$ and can thus be neglected.

The most up-to-date lattice QCD results for the magnetic susceptibility can 
be found in Ref.~\cite{Bali:2014kia}. A simple parameterization 
that agrees with the data within two standard deviations 
above $T=110\textmd{ MeV}$ and matches 
perturbation theory at high temperatures (cf.\ Ref.~\cite{Bali:2014kia}) is
\begin{equation}
T>110 \textmd{ MeV}: \quad 
\chi(T) = \frac{e^2}{3\pi^{2}}\log\frac{T}{110 \textmd{ MeV}},
\label{eq:susc}
\end{equation}
where $e$ denotes the elementary charge. Below it will be convenient to 
give the magnetic field in terms of $e$, since the combination 
$eB$ has units $\textmd{GeV}^2$. 
Notice that we use isothermal freezeout conditions, 
where the hypersurface is determined by
a constant temperature $T=137$ MeV.

\subsection*{Magnetic field profile}

In order to calculate $F^\nu$, knowledge of the spatial profile of the magnetic field 
is necessary at each point in time. As it turns out, the largest uncertainty in 
this description is the time-dependence $\mathbf{B}(\tau)$.
While the magnetic field due to the spectators would drop very quickly 
in the vacuum~\cite{Kharzeev:2007jp} (in fact, by a few orders of magnitude within $1 \textmd{ fm}/c$), it has been 
speculated that the field could survive much longer in the QGP 
if the electrical conductivity of the plasma is 
high~\cite{McLerran:2013hla,Tuchin:2013ie,Li:2016tel}.
In the pre-equilibrium stage, the decay of the magnetic field might also 
be delayed by charged quark-antiquark pairs
due to gluon splitting and the Schwinger mechanism~\cite{Schwinger:1951nm}, 
see also Ref.~\cite{Tuchin:2013ie}.
In this paper, the terms `medium' and `vacuum' are used to denote a collision system with or without electrical conductivity.

Although the evolution of the magnetic field is likely to be 
affected considerably by the medium, the dependence 
of $B$ on the collision parameters can be estimated by using 
the Lienard-Wiechert potential of the colliding nucleons in the vacuum.
While the magnetic field was found~\cite{Skokov:2009qp} 
to depend rather weakly on the electric charge 
number $Z$ of the nuclei ($B\propto Z^{1/3}$), 
it was observed to be strongly influenced by the 
impact parameter $b$ and the collision beam energy 
$\sqrt{s_{NN}}$~\cite{Bzdak:2011yy,Deng:2012pc,Kharzeev:2007jp}.
The spatial distribution of the magnetic field in the reaction plane 
was studied both in a setting with uniform nucleon density in the 
nucleus~\cite{Tuchin:2013apa} as well as a more realistic 
Woods-Saxon nucleon density
distribution~\cite{Deng:2012pc,Zhong:2014sua},
showing a pronounced anisotropy in the magnetic field profile -- namely 
a steep fall-off along the impact parameter direction and a slower 
variation perpendicular to the reaction plane. 

Based on these considerations, the transverse distribution of a longitudinal boost invariant -- i.e., $\eta_s$-independendent -- magnetic field is parameterized as,
\begin{equation}
eB(x,y,\tau)=eB_{0}\exp\left(-\frac{x^{2}}{2\sigma_{x}^{2}}-\frac{y^{2}}{2\sigma_{y}^{2}}\right) \exp\left(-\frac{\tau}{t_{d}}\right),\label{eq:magnetic_eB}
\end{equation}
where $eB_{0}$ is the amplitude of the magnetic field, which is
taken to be $0.09\ \mathrm{GeV^{2}}$ ($\approx 5 \,m_{\pi}^2$) for RHIC energy
and $1.33\ \mathrm{GeV^{2}}$ ($\approx 70\,m_{\pi}^2$) for LHC energy~\cite{Deng:2012pc} while $\sigma_{x}$ and $\sigma_{y}$
are the Gaussian widths along the $x$ and $y$ directions, respectively.
Since the time evolution of the magnetic field in the pre-equilibrium stage is still unknown,
we use an exponential decay for this, with a lifetime $t_{d}$ varying between
$0.1\textmd{ fm}$ and $1.9 \textmd{ fm}$. 
The various different settings used in the paper are summarized in Table.~\ref{tab:em_settings}.
\begin{table}[h]
  \centering
\begin{tabular}{|l|c|c|c|c|c|}
  \hline
  setting  & $eB_0\ [\textmd{GeV}^2]$ & $t_d\ [\textmd{fm}]$ & $\sigma_x\ [\textmd{fm}]$ & $\sigma_y\ [\textmd{fm}]$ \\
  \hline
  A       &  0.09 & 1.9   & 1.3        & 2.6      \\
  \hline
  B       &  1.33 & 0.1   & 1.3        & 2.6      \\
  \hline
  C       &  1.33 & 0.5   & 1.3        & 2.6      \\
  \hline
  D       &  1.33 & 1.0   & 1.3        & 2.6      \\
  \hline
  E       &  1.33 & 1.9   & 1.3        & 2.6      \\
  \hline
  F       &  1.33 & 1.9   & 2.4        & 4.8      \\
  \hline
  G       &  1.0 & 1.9   & 2.4        & 4.8      \\
  \hline
\end{tabular}
\caption{\label{tab:em_settings} The configurations for the space-time profile of the magnetic field,
inspired by previous studies of magnetic fields in the vacuum~\cite{Kharzeev:2007jp,Zhong:2014sua}
and in the QGP \cite{Gursoy:2014aka,Zakharov:2014dia,Tuchin:2013apa,Gupta:2003zh,Qin:2013aaa,Greif:2014oia,McLerran:2013hla,Tuchin:2013ie} (settings A-G)
together with the reference configuration at $eB=0$.
}
\end{table}

\begin{figure}[!htp]
  \includegraphics[width=0.5\textwidth, height=0.4\textwidth]{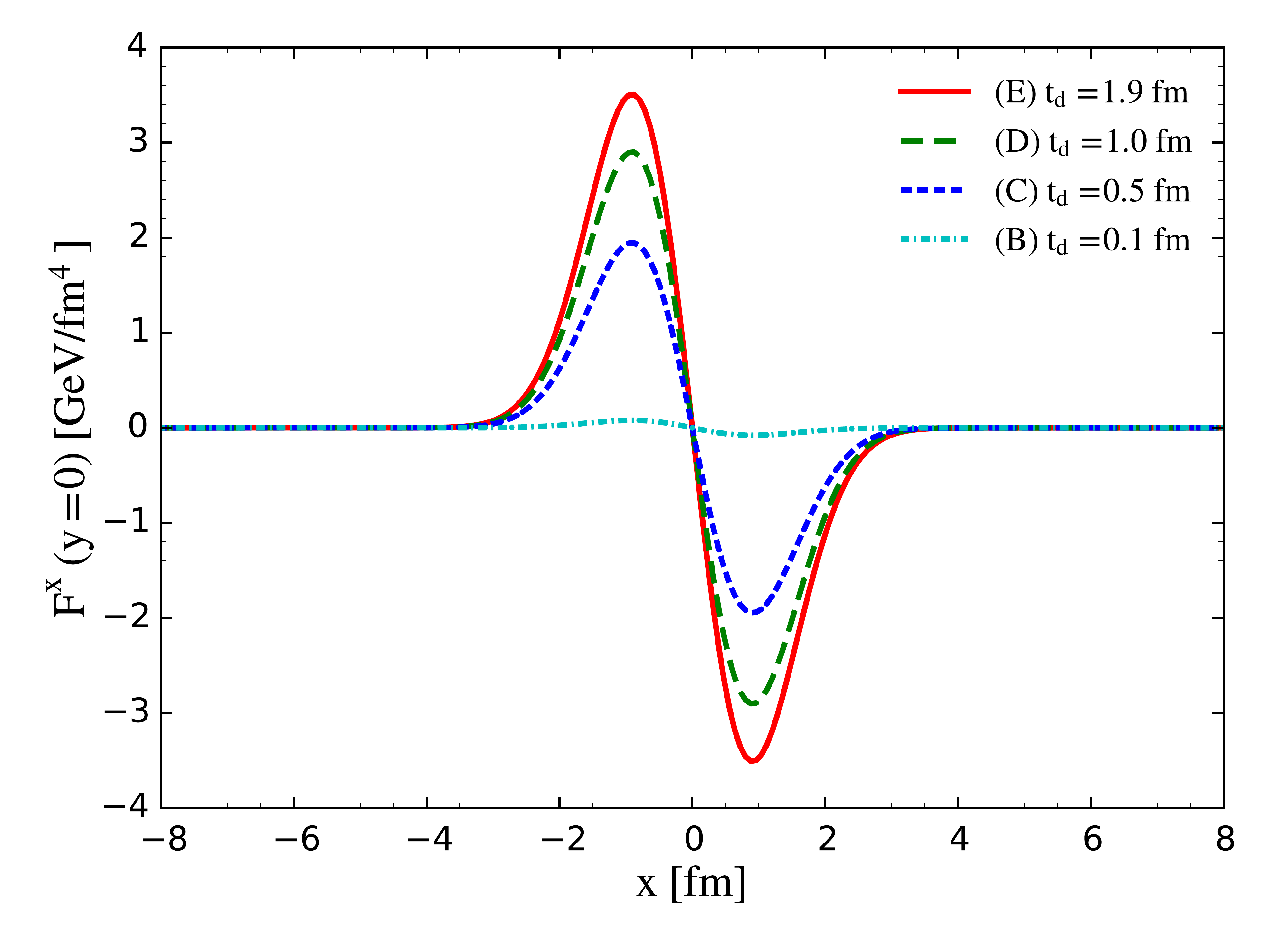}
  \caption{The squeezing force density ($F^{x}$) along the $x$ axis at $\tau_0=0.2$ fm with
    the magnetic field given by settings (B) $t_d=0.1$ fm, (C) $t_d=0.5$ fm, (D) $t_d=1.0$ fm and (E) $t_d=1.9$ fm,
    for Pb+Pb $\sqrt{s_{NN}}=2.76$ TeV collisions with local temperature from (3+1)D hydrodynamics.
  \label{fig:force_density_CDEF_LHC}}
\end{figure}

Four different parameters for the lifetime $0.1\textmd{ fm}\le t_d\le 1.9 \textmd{ fm}$
are used for Pb+Pb collision to get the spatial distribution of the squeezing force density during the hydrodynamic evolution (which starts at the 
thermalization time $\tau_0$).
As shown in Fig.~\ref{fig:force_density_CDEF_LHC}, the maximum force density 
at time $\tau_0=0.2$ fm is in the range $(2\ldots4)\ \mathrm{GeV/fm^4}$ 
for long-lived fields, 
whereas it is merely $0.08\ \mathrm{GeV/fm^4}$ for $t_d=0.1$ fm.

\subsection*{Initial state and thermalization time}

The initial energy density distribution is given by the optical Glauber model, %in Eq.~\ref{eq:ini_ed},
\begin{equation}
 \begin{split}
  \varepsilon(\tau_0, x&, y, \eta_s) = K\cdot (0.95N_{wn}+0.05N_{bc}) \\
  &\times \exp\left[-\frac{(|\eta_s|-\eta_{w}/2)^2}{2\sigma_{\eta}^2}\,\Theta(|\eta_s|-\eta_{w}/2) \right],
 \end{split}
 \label{eq:ini_ed}
\end{equation}
where a large fraction of the initial energy deposition is assumed to come from the soft part that is proportional to $0.95N_{wn}$
($N_{wn}$ is the number of wounded nucleons), and a smaller fraction %of the initial energy deposition 
is proportional to $0.05N_{bc}$
($N_{bc}$ is the number of binary collisions).
In calculating $N_{wn}$ and $N_{bc}$, the inelastic scattering cross sections $\sigma_0$ 
are set to $64$ mb for Pb+Pb $2.76$ TeV and $40$ mb for Au+Au $200$ GeV collisions.
One envelope distribution is used along the longitudinal direction, where the width of the plateau $\eta_{w}$ at mid-rapidity
and the width of the fast fall-off $\sigma_{\eta}$ at large rapidity are set to $5.9$ and $0.4$ for Au+Au
$\sqrt{s_{NN}}=200$ GeV and $7.0$ and $0.6$ for Pb+Pb $\sqrt{s_{NN}}=2.76$ TeV collisions.
The parameter $K$ is fixed by the maximum energy density $\varepsilon_0$ given in 
Table.~\ref{tab:edmax},
for most central collisions at RHIC and LHC energy.

In the optical Glauber model, the nucleon densities of the Pb and Au nucleus are described by the Woods-Saxon distribution,
\begin{equation}
  \rho(r) = \frac{\rho_0}{\exp\left( \frac{r-R}{d} \right)+1},
  \label{eq:woods_saxon}
\end{equation}
where R is the radius of the nucleus, $\rho_0$ is the average nucleon density and $d$ is the diffusiveness.
The parameters used in the optical Glauber model are listed in Table.~\ref{tab:glauber},
\begin{table}[h]
  \centering
\begin{tabular}{|l|c|c|c|c|}
  \hline
  nucleus & A & $\rho_0$ $[1/\textmd{fm}^3]$  & R $[\textmd{fm}]$  & d $[\textmd{fm}]$\\
  \hline
  Pb      & 208 & $0.17$    & 6.38 & 0.535  \\
  \hline
  Au      & 197 & $0.17$    & 6.62 & 0.546  \\
  \hline
\end{tabular}
\caption{\label{tab:glauber} Parameters used in the Woods-Saxon distribution for Pb and Au nucleus.}
\end{table}

\begin{figure}[t]
\includegraphics[width=0.5\textwidth]{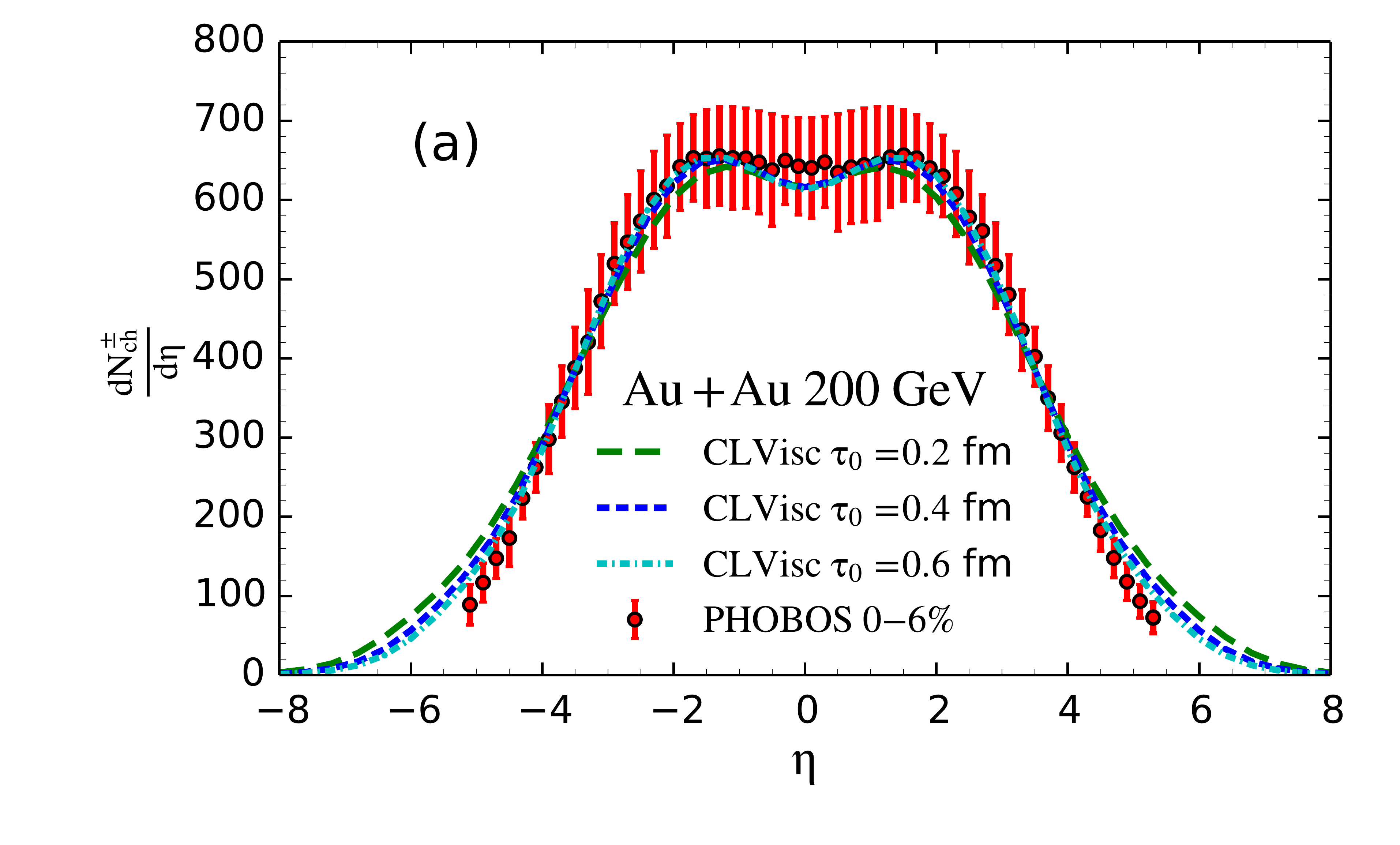}
\includegraphics[width=0.5\textwidth]{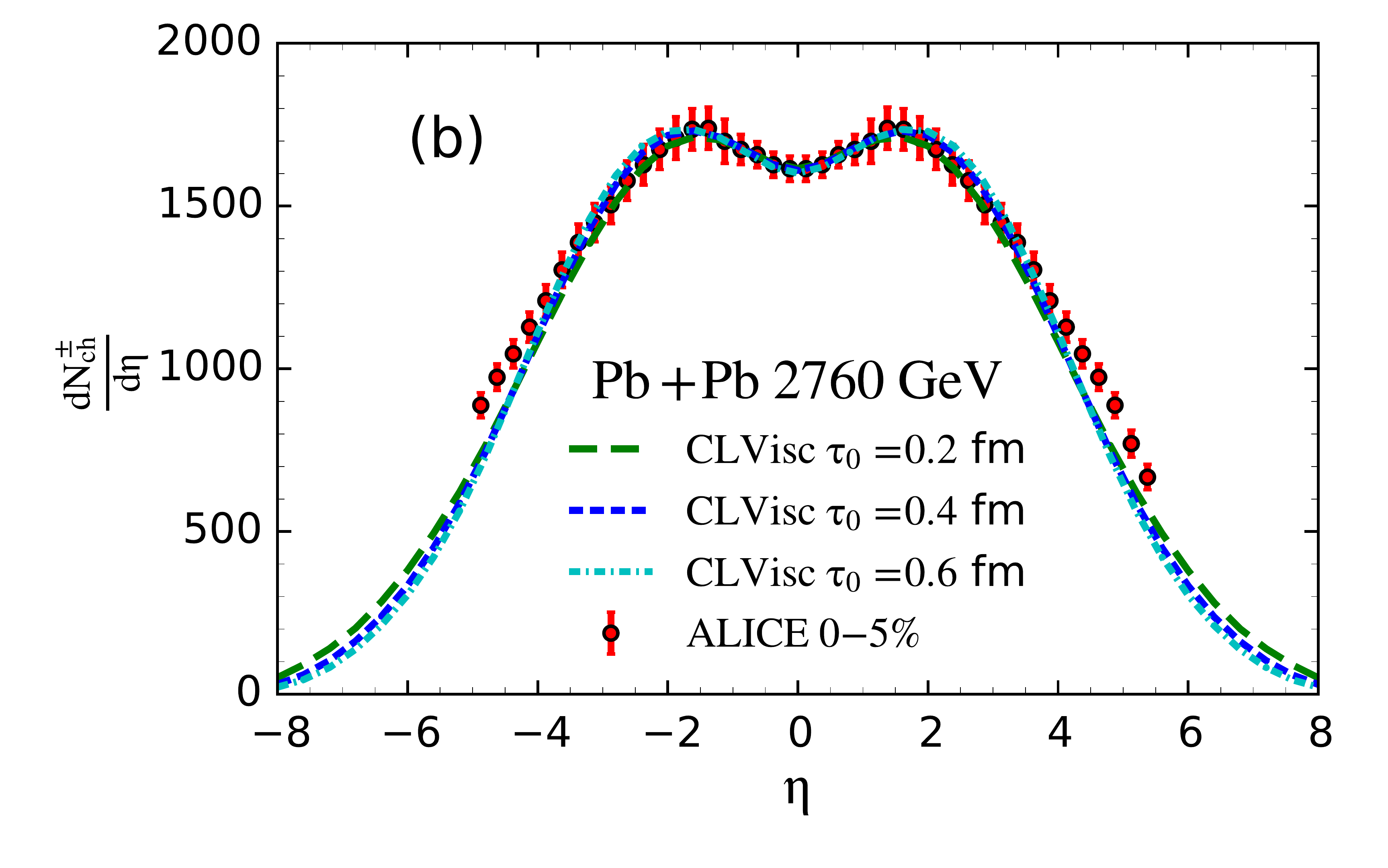}
  \caption{(color online) The charged multiplicity for (a) Au+Au $\sqrt{s_{NN}}=200$ GeV collisions and (b) Pb+Pb 
  $\sqrt{s_{NN}}=2.76$ TeV collisions with 3 different groups of configurations for initial thermalization time $\tau_0$
  and maximum energy density $\varepsilon_0$ given in Table. \ref{tab:edmax}.
  \label{fig:multiplicity}}
\end{figure}

The effect of the squeezing force (with small $t_d$) is sensitive to the strength of the magnetic field at initial thermalization time $\tau_0$.
According to the analytical solution of 1D Bjorken hydrodynamics 
$$ \varepsilon/\varepsilon_0 = (\tau_0/\tau)^{1+c_s^2}, \quad\; s/s_0=(\varepsilon/\varepsilon_0)^{1/(1+c_s^2)},$$
where $\varepsilon$($\varepsilon_0$) and $s$($s_0$) are the energy density and entropy density at $\tau$($\tau_0$) respectively and $c_s$ is the speed of sound.
Therefore, in this approximation the evolution $s(\tau)$ is invariant under changes
of $\tau_0$ as long as the combination $\tau_0 \mathrm{s}_0$ is kept constant.
(The entropy density at $\tau_0$ is obtained from the EoS and the 
maximum energy density $\epsilon_0$.)
As shown in Table.~\ref{tab:edmax}, three groups of initial thermalization time $\tau_0$ and maximum energy density $\varepsilon_0$
are listed that give the same charged multiplicity for most central collisions with 
mean impact parameter $\langle b \rangle=2.4$ fm for RHIC and $\langle b \rangle=2.65$ fm for LHC energy as shown in Fig.~\ref{fig:multiplicity}.
Notice that the initial settings $(\tau_0=0.4\ \textmd{fm}$, $\varepsilon_0=55\ \textmd{GeV})$ are used in
\cite{Schenke:2010nt} to fit charged multiplicity for Au+Au $\sqrt{s_{NN}}=200$ GeV collisions in centrality class $0-6\%$,
which is a cross check for the current calculation.
We will get back to the $\tau_0$-dependence of the squeezing effect below in 
Sec.~\ref{sec:results}.
\begin{table}[h]
  \centering
\begin{tabular}{|l|c|c|c|}
  \hline
  $\tau_0\ [\mathrm{fm}]$  & 0.2  & 0.4 & 0.6 \\
  \hline
  $\varepsilon_0 [\mathrm{GeV/fm^3}]$ at RHIC & 135.5 & 55.0 & 32.6 \\
  \hline
  $\varepsilon_0 [\mathrm{GeV/fm^3}]$ at LHC & 413.9 & 166.4 & 98.0 \\
  \hline
\end{tabular}
\caption{\label{tab:edmax}Maximum energy density for (3+1)D ideal hydrodynamics starting from different values of $\tau_0$
  to get the same charged multiplicity distribution for RHIC and LHC energy.}
\end{table}

\subsection*{Quantities of Interest/Observables}

Since in non-central collisions the pressure gradients are enhanced along the $x$ direction (transverse direction in reaction plane) and suppressed along the 
$y$ direction (perpendicular to reaction plane), the 
fireball expands faster along $x$ for such events. Accordingly, the 
final charged hadrons from the collective expansion
have larger transverse momentum along this direction.
How the spatial eccentricity $\epsilon_{x}$ is
transferred to the momentum anisotropy $\epsilon_{p}$ is described step-by-step in the hydrodynamic
simulations. The quantities $\epsilon_{x}$ and $\epsilon_{p}$ are useful to 
quantify the effect of the squeezing
force density on the anisotropic expansion. To be precise, $\epsilon_{x}$ and $\epsilon_{p}$ are defined
as,
\begin{eqnarray}
\epsilon_{x} & = & \frac{\langle\varepsilon\gamma(y^{2}-x^{2})\rangle}{\langle\varepsilon\gamma(y^{2}+x^{2})\rangle},\label{eq:ecc_x}\\
\epsilon_{p} & = & \frac{\langle T^{xx}-T^{yy}\rangle}{\langle T^{xx}+T^{yy}\rangle},\label{eq:ecc_p}
\end{eqnarray}
where $\langle \rangle$ is an average over the transverse plane, $\varepsilon$
and $\gamma$ are the energy density and Lorentz factor, respectively, and
$T^{xx}$ and $T^{yy}$ are the two diagonal components of the energy momentum
tensor.

Another quantity that reflects the momentum anisotropy is the $p_T$ differential elliptic flow $v_2$
of final charged hadrons at mid-rapidity, which is defined as,
\begin{equation}
  v_2(p_T) \equiv \frac{\int d\phi \frac{dN}{dYdp_Td\phi} \cos(2(\phi-\Psi_2))}{\int d\phi \frac{dN}{dYdp_Td\phi}}.
  \label{eq:v2}
\end{equation}
Here, $\Psi_2$ is the event plane, which equals zero in the current study neglecting event-by-event fluctuations.

\section{Results \label{sec:results}}

\subsection*{\boldmath Pb+Pb $\sqrt{s_{NN}}=2.76$ TeV collisions}

\begin{figure}[t]
  \includegraphics[width=0.5\textwidth]{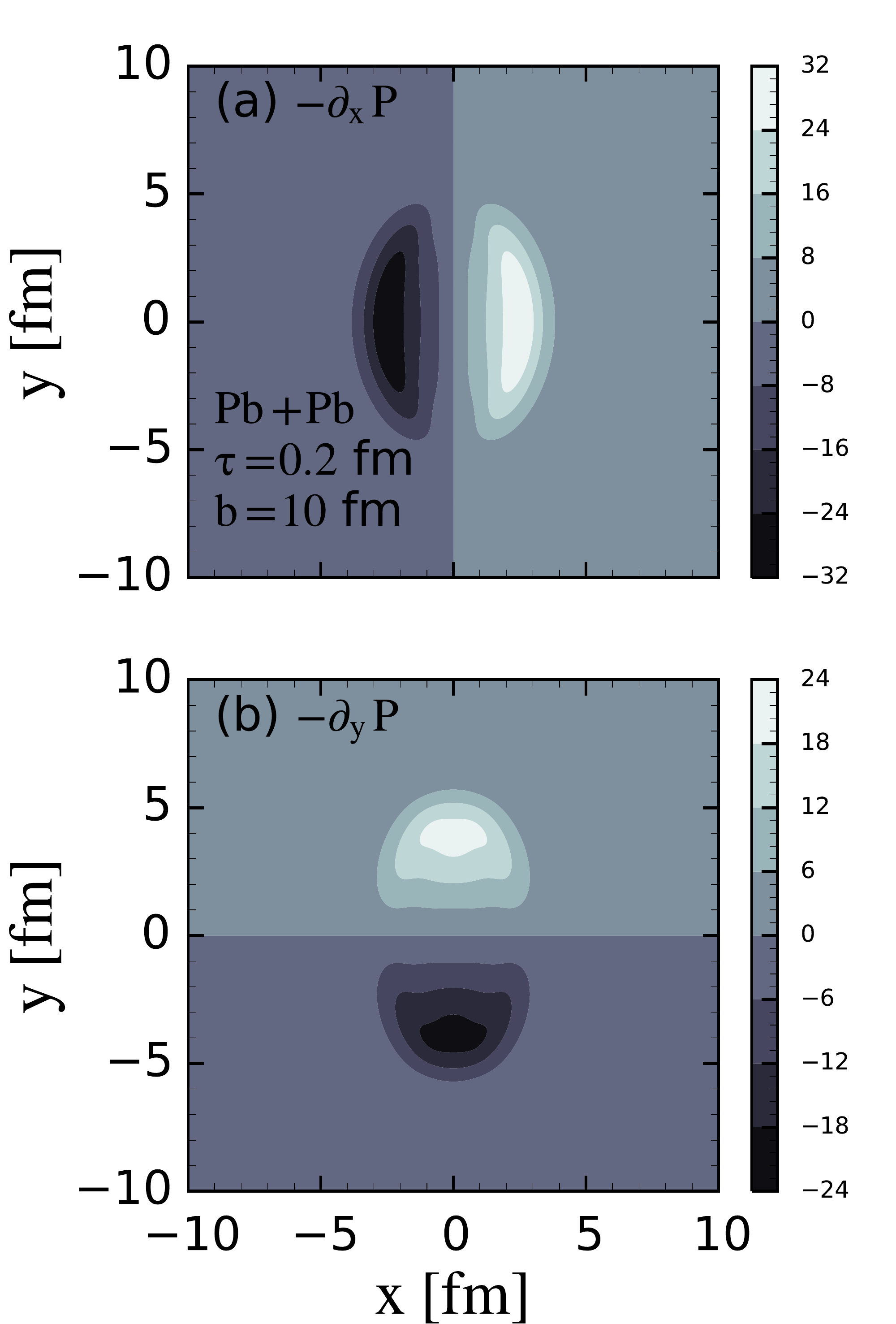}
  \caption{The pressure gradients for Pb+Pb $\sqrt{s_{NN}}=2.76$ TeV collisions with impact parameter $b=10$ fm.
  \label{fig:pressure_gradient_LHC}}
\end{figure}

Before performing full calculations, a simple comparison between the pressure gradients and the squeezing force 
density due to the magnetic field at thermalization time $\tau_0$ is helpful to provide intuitive estimates. For the following results we set $\tau_0=0.2$ fm/c.
The initial pressure gradients in the transverse plane (mid-rapidity) for Pb+Pb $\sqrt{s_{NN}}=2.76$ TeV
collisions with impact parameter $b=10$ fm are shown in Fig.~\ref{fig:pressure_gradient_LHC}. 
As expected, the pressure gradient along the $x$ direction exceeds that along the 
$y$ direction, with a maximal value of around $32$ $\mathrm{GeV/fm^4}$. 

\begin{figure}[!htp]
  \includegraphics[width=0.5\textwidth]{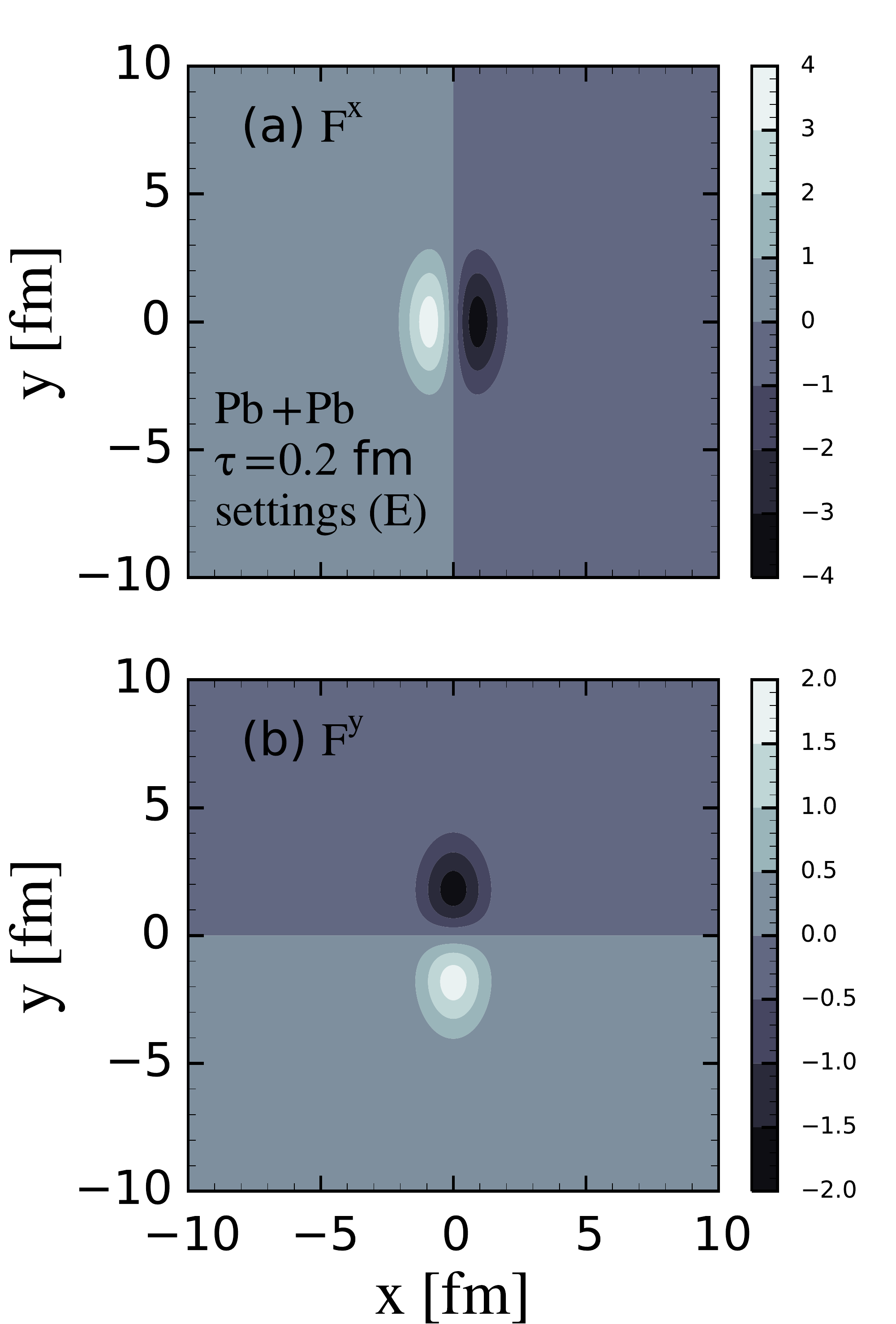}
  \caption{The squeezing force density (a) along the $x$ direction and (b) along 
  the $y$ direction, 
    for Pb+Pb $\sqrt{s_{NN}}=2.76$ TeV collisions at $\tau_0=0.2$ fm, with
    the magnetic field given by setting (E).
  \label{fig:force_density_LHC}}
\end{figure}

For a first calculation of the squeezing force density, 
we take the magnetic field profile of setting (E), which 
provides the highest initial magnetic field $B_0$ and the longest lifetime $t_d$. 
As Fig.~\ref{fig:force_density_LHC} shows, 
the force density distribution\footnote{Note that the local temperature given by the hydrodynamic simulation 
is employed in the susceptibility~(\ref{eq:susc}) to 
calculate the squeezing force density.} is of similar shape
and has magnitudes of about $10\%$ compared to the pressure gradients.
We also observe that the direction of the force density is roughly 
opposite to the pressure gradients -- indeed providing a squeezing effect.
For setting (E), where $2\sigma_x=\sigma_y=2.6$ fm, the maximum squeezing force density is located at $x=\pm 1$ fm,
while the maximum pressure gradient at $x=\pm 2$ fm.
The region with high pressure gradients is more extended than the 
region, where the force density is pronounced. 
Thus it is also of interest to resize the high squeezing force density region by varying $\sigma_x$ and $\sigma_y$,
to investigate the effect on the anisotropic flow (setting (F)).

\begin{figure}[!htp]
\includegraphics[width=0.5\textwidth]{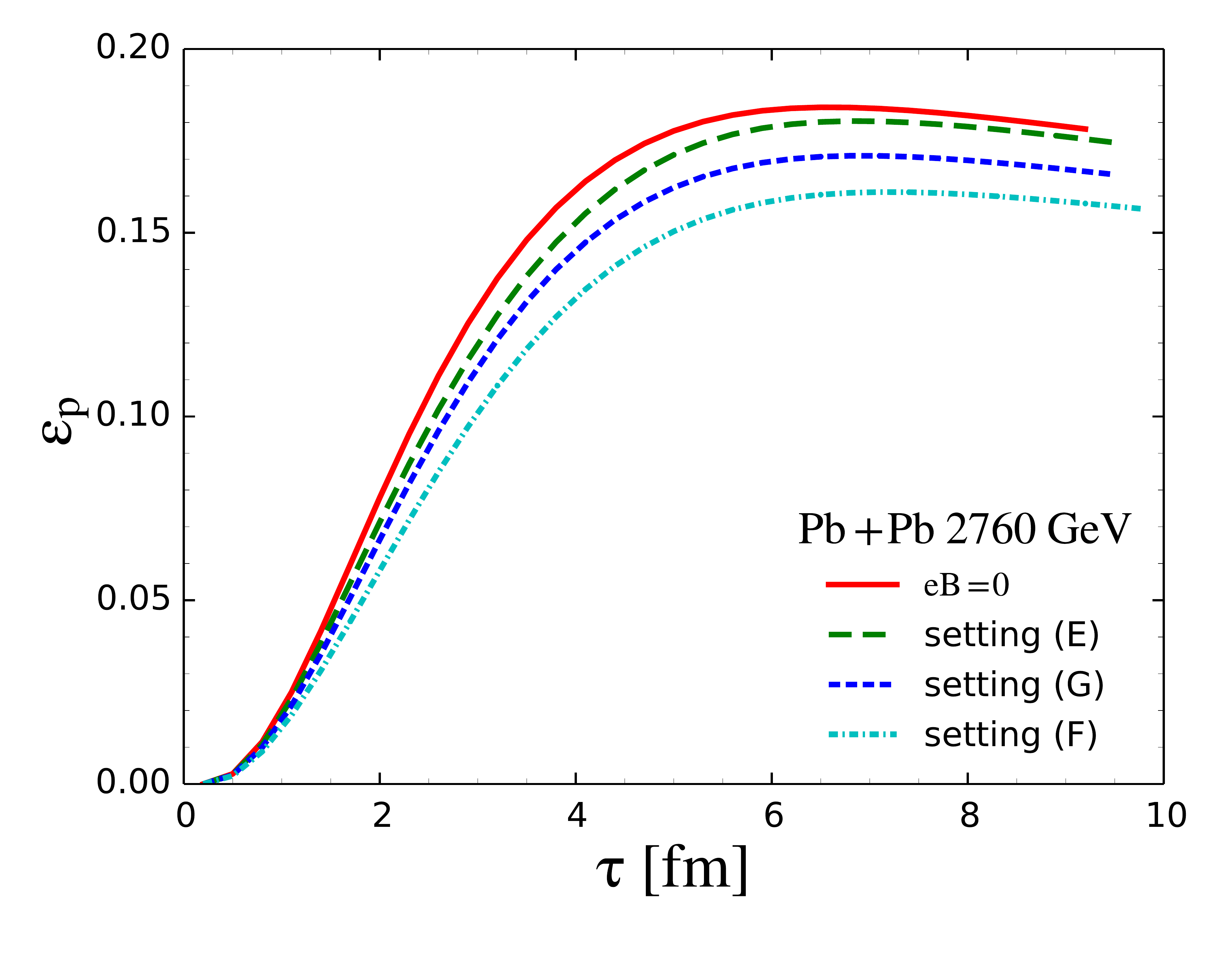}
  \caption{(color online) Time evolution of the momentum eccentricity $\epsilon_p$ for Pb+Pb $\sqrt{s_{NN}}=2.76$ TeV collisions with impact parameter $b=10$ fm, for 
    (solid-line) $eB_0=0$, setting
    (E) (long-dashed line, $eB_0=1.33\ \mathrm{GeV^2}$, $2\sigma_x=\sigma_y=2.6\ \textmd{fm}$), 
    setting (G) (short-dashed line, $eB_0=1.0\ \mathrm{GeV^2}$, $2\sigma_x=\sigma_y=4.8\ \textmd{fm}$)
    and setting (F) (dash-dotted line, $eB_0=1.33\ \mathrm{GeV^2}$, $2\sigma_x=\sigma_y=4.8\ \textmd{fm}$).
\label{fig:ecc_LHC}}
\end{figure}

Since the spatial distributions of both the squeezing force density and of the pressure gradients evolve with time,
the previous comparison at $\tau=\tau_0$ clearly does not capture the full effect.
To take into account the competition between the squeezing force density and
the pressure gradients at each time step, we add the former 
as a source term for the energy-momentum tensor in the hydrodynamic 
equations, as in Eq.~(\ref{eq:magnetic_fluid}).
In Fig.~\ref{fig:ecc_LHC} we show the time evolution of the momentum anisotropy $\epsilon_p$
in the 
transverse plane for $B=0$ and for settings (E), (G) and (F), cf.\ Tab.~\ref{tab:em_settings}. 
The squeezing force density with setting (E) reduces the momentum anisotropy by $5\%$ at intermediate time, while 
that with larger $\sigma_x$ and $\sigma_y$ in setting (F) reduces $\epsilon_p$ 
by as much as $20\%$.
Notice that the lifetime of the fireball in all cases increases, because the magnetic 
forces compress the system and, thus, reduce the rate of expansion.

\begin{figure}[!htp]
\includegraphics[width=0.5\textwidth]{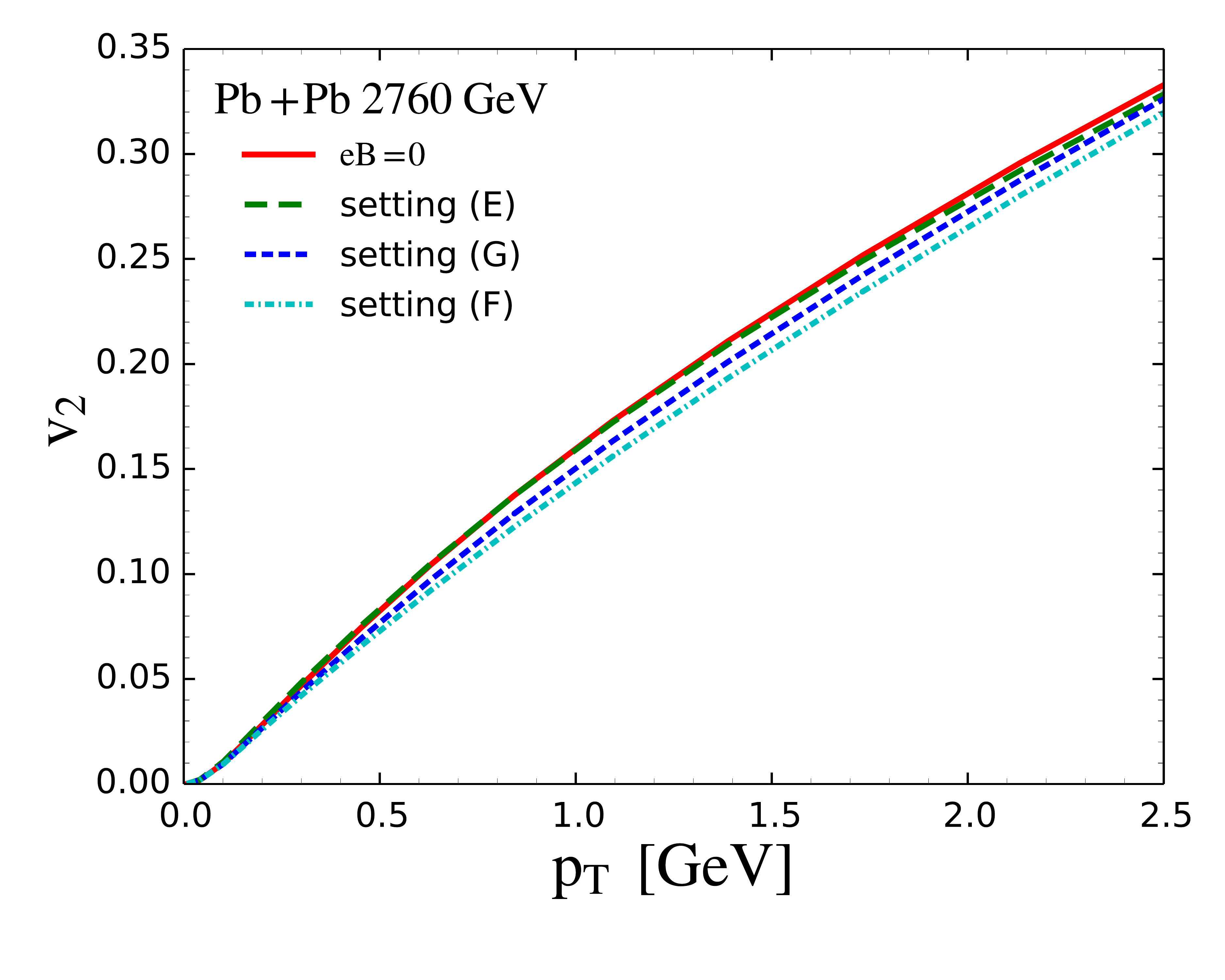}
  \caption{(color online) $v_2$ for direct $\pi^+$ at Pb+Pb $\sqrt{s_{NN}}=2.76$ TeV collisions with impact parameter $b=10$ fm,
for (solid-line) $eB_0=0$, settings (E) (long-dashed line) $eB_0=1.33\ \mathrm{GeV^2}$, $2\sigma_x=\sigma_y=2.6\ \textmd{fm}$,
(G) (short-dashed line) $eB_0=1.0\ \mathrm{GeV^2}$, $2\sigma_x=\sigma_y=4.8\ \textmd{fm}$ and
(F) (dash-dotted line) $eB_0=1.33\ \mathrm{GeV^2}$, $2\sigma_x=\sigma_y=4.8\ \textmd{fm}$.
  \label{fig:v2_LHC}}
\end{figure}

The impact on the $p_T$ differential anisotropic flow for direct $\pi^+$ emitted from the freezeout hypersurface, on the other hand,
is less dramatic, as shown in Fig.~\ref{fig:v2_LHC}. The squeezing force density with setting (E) reduces the elliptic flow by merely $1-2\%$. Increasing the 
spatial size of the magnetized region (setting (F)), 
the suppression of $v_2$ becomes around $6\%$. 
Both for $\epsilon_p$ and for $v_2$, effects of similar size are observed for 
weaker but wide-spread magnetic fields, as given in setting (G) $eB_0 \approx 50\, m_\pi^2$.

We also remark that besides $B_0$, $t_d$ and $\sigma_{x,y}$, a similarly relevant 
role is played by the thermalization time $\tau_0$. A lower $\tau_0$ enhances 
the squeezing force in two ways. First, because the magnetic field is higher 
at earlier times and, thus, the gradient of $\mathbf{B}$ is also increased. 
Second, because the temperature is also higher and the hot QGP is more prone 
to the squeezing effect due to its enhanced magnetic susceptibility, see Eq.~(\ref{eq:susc}).
At the same time, the initial energy density and, thus, the pressure gradients are also very high for small $\tau_0$.
The net dependence of the effect on $\tau_0$ could thus be
non-trivial.

\begin{figure}[!htp]
  \includegraphics[width=0.5\textwidth]{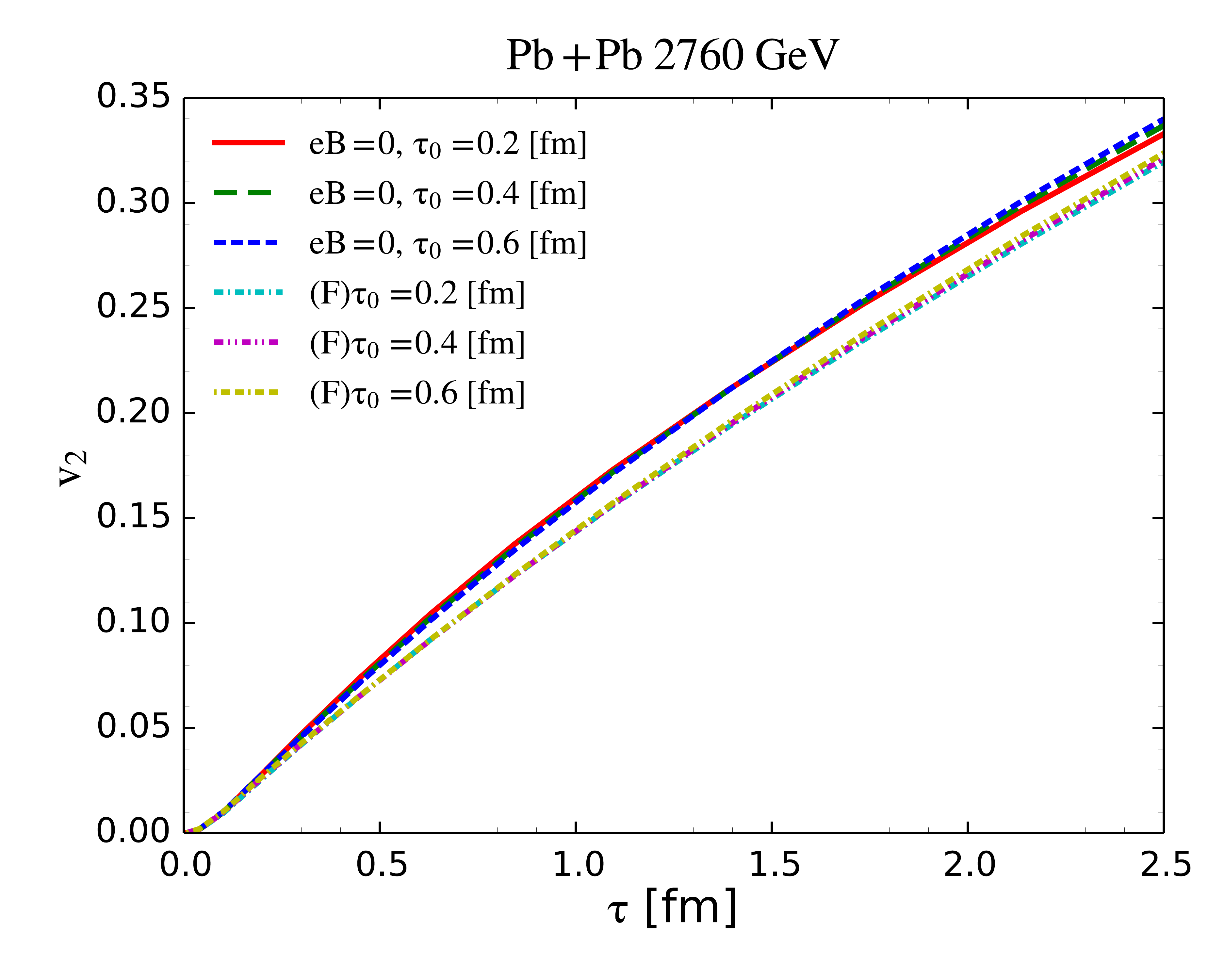}
  \caption{(color online) The effect of paramagnetic squeezing on the elliptic flow for $\tau_0=0.2,\ 0.4\ \textmd{and}\ 0.6$ fm.
  \label{fig:tau0_dep}}
\end{figure}

In general, starting the hydrodynamic evolution at a larger value of $\tau_0$ shifts the 
$\tau$-dependence of $\epsilon_p$ to later times due to the zero initial transverse flow assumed in this study.
However, as shown in Fig.~\ref{fig:tau0_dep}, the elliptic flow on the freezeout hypersurface does not change considerably for three different values of $\tau_0$
in the absence of magnetic fields.
For a long-lived magnetic field with $t_d \gg \tau_0$,
the squeezing effect on $v_2$ is also insensitive to $\tau_0$.
Note however, that this would not be the case for a magnetic field of shorter 
lifetime $t_d\approx \tau_0$. In the latter range of proper times, the exponential decay $B(\tau)$ suppresses the magnetic field -- and, thus, the effect -- considerably.

\subsection*{\boldmath Au+Au $\sqrt{s_{NN}}=200$ GeV collisions}

\begin{figure}[!htp]
  \includegraphics[width=0.5\textwidth]{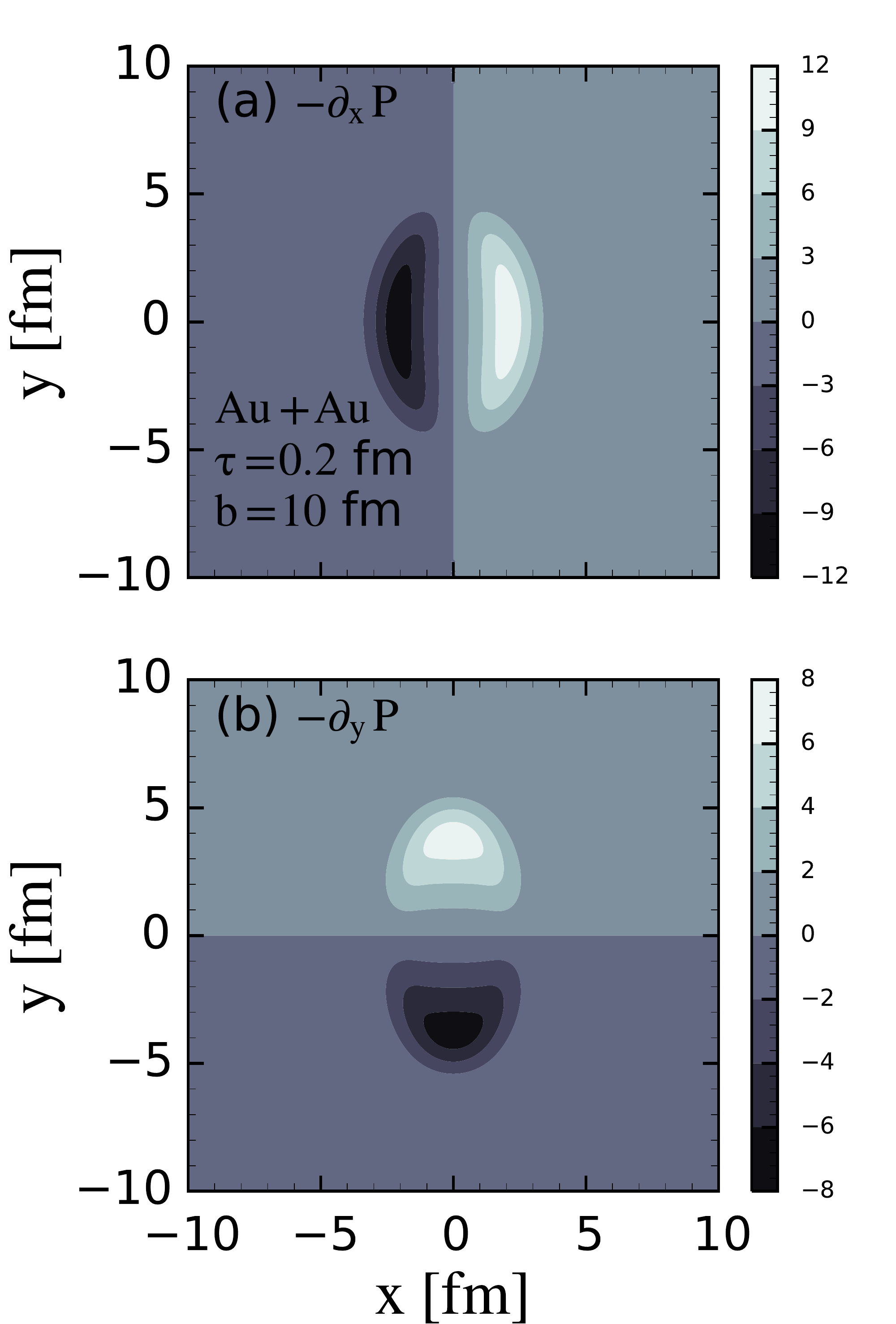}
  \caption{The pressure gradients for Au+Au $\sqrt{s_{NN}}=200$ GeV collisions with impact parameter $b=10$ fm.
  \label{fig:pressure_gradient_RHIC}}
\end{figure}
\begin{figure}[!htp]
  \includegraphics[width=0.5\textwidth]{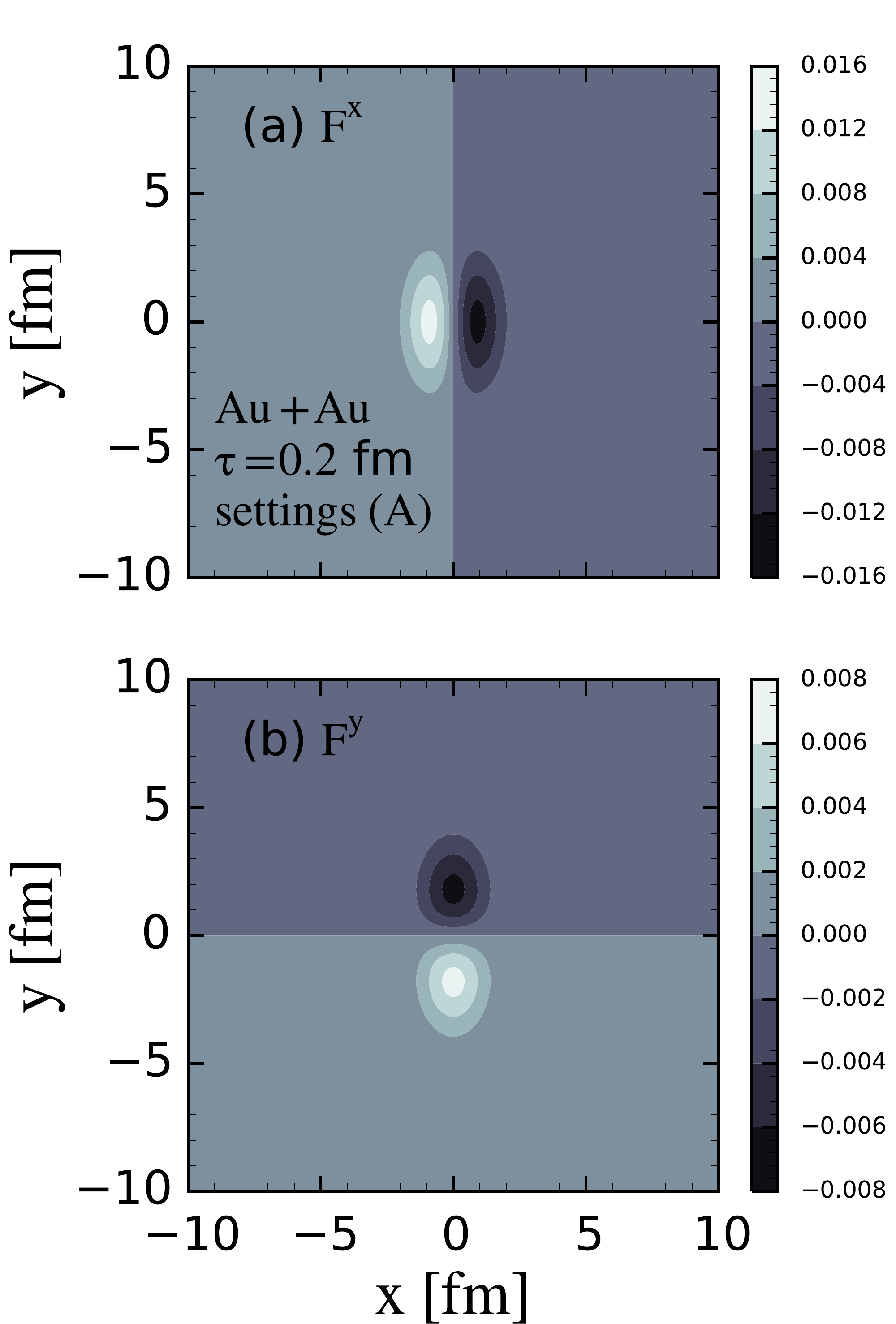}
  \caption{The squeezing force density for Au+Au $\sqrt{s_{NN}}=200$ GeV collisions at $\tau_0=0.2$ fm, with
    the magnetic field given by setting (A).
  \label{fig:squeezing_RHIC}}
\end{figure}
Previous studies have shown that the magnetic
field at RHIC energy decays much slower in the vacuum than at LHC energy due to the smaller
relative speed between the colliding nuclei~\cite{Zhong:2014sua}.
Here, we use a long-lived magnetic field with lifetime $t_d=1.9$ fm at RHIC energy
(setting (A)) to estimate the magnitude of the squeezing force density at $\tau_0=0.2$ fm.
The maximum magnetic field at RHIC energy ($0.09$ $\mathrm{GeV^{2}}$
used in this paper) is much smaller than that at LHC energy ($1.33$  $\mathrm{GeV^{2}}$).
The ratio of the maximal squeezing force density between RHIC and LHC collisions
is thus proportional to $(0.09/1.33)^2\approx0.0046$.
On the other hand, the maximal pressure gradients in the QGP at RHIC energy is
around $1/3$ of that at LHC energy, compare Fig.~\ref{fig:pressure_gradient_RHIC}
to Fig.~\ref{fig:pressure_gradient_LHC}.
The distribution of the squeezing force density for RHIC collisions 
is shown in Fig.~\ref{fig:squeezing_RHIC}, revealing that the paramagnetic squeezing
is completely negligible at RHIC energy in the current framework.

\section{Summary}
\label{sec:summary}

We have studied the effect of paramagnetic squeezing~\cite{Bali:2013owa} on the anisotropic expansion
of bulk matter in non-central high-energy heavy-ion collisions, within a (3+1)-dimensional ideal hydrodynamic model.
We found that the effect is sensitive to the space-time distribution of the magnetic field.
A sizable effect is observed for long-lived and wide-spread magnetic fields at LHC energy.
For such magnetic fields, the lifetime of the fireball (which is defined as the time interval between $\tau_0$ and
the complete freeze out when the temperature of all the fluid cells below $T_{frz}$)
is elongated by $\sim 5\%$ as shown in Fig.~\ref{fig:ecc_LHC},
the momentum eccentricity averaged on the whole bulk matter reduces by $10-20\%$,
while the anisotropic flow of $\pi^+$ from the freezeout hypersurface reduces by $6\%$.
The slight reduction of $v_2$ due to the effect implies that the elliptic flow 
measured in the experiment is somewhat smaller than what would result from only the 
viscous flow of the plasma. In other words, interpreting the experimental results 
for $v_2$ without taking the squeezing effect into account slightly 
overestimates the ratio $\eta/s$ of the QGP.
Strong suppression of momentum anisotropy $\epsilon_p$ also
suggests that electromagnetic probes such as thermal photons and di-leptons might be
more sensitive to the paramagnetic squeezing,
since they are emitted from the whole QGP and HRG stage of the evolution and not just from the freezeout hypersurface.
In addition, we also observed that the effect of the squeezing force density is negligible at RHIC energy,
because the squeezing force density decreases much faster than the pressure gradients
as the beam energy is reduced.

The current work can be extended in several ways.
First, in real heavy-ion collisions, the initial charge density fluctuates strongly and the 
initial energy density is very lumpy. It is possible that the local magnetic field is large
while the energy density is small~\cite{Roy:2015kma}. The ratio between the squeezing force density and the
pressure gradients for those regions can be very large and the effect for the anisotropic flow
might be different.
Second, the transport coefficients of the QGP are affected by strong magnetic fields:
the shear viscosity $\eta$ in the direction perpendicular to the magnetic field
may be twice as small as in the direction of the field~\cite{Tuchin:2011jw,Tuchin:2013ie,Critelli:2014kra}. 
The elliptic flow of final charged hadrons is predicted to be affected by
this asymmetric shear viscosity over entropy density ratio $\eta/s$ according to early studies
in 2nd order viscous hydrodynamics \cite{Romatschke:2007mq,Song:2007fn}.
Relativistic magnetohydrodynamics with anisotropic shear viscosity needs to be developed for more accurate
studies in the future.

\begin{acknowledgments}
  We thank Andreas Sch\"{a}fer for helpful discussions. LG.\ Pang and H.\ Petersen acknowledge funding
  of a Helmholtz Young Investigator Group VH-NG-822 from the Helmholtz
  Association and GSI. 
  G.\ Endr\H{o}di acknowledges support from the DFG (Emmy Noether 
  Programme EN 1064/2-1; and SFB/TRR 55).
  This work was supported in part by the Helmholtz International
  Center for the Facility for Antiproton and Ion Research (HIC for FAIR)
  within the framework of the Landes-Offensive zur Entwicklung Wissenschaftlich-Oekonomischer
  Exzellenz (LOEWE) program launched by the State of Hesse, 
  Computational resources have been provided by the Center for Scientific Computing
  (CSC) at the Goethe-University of Frankfurt.

\bibliographystyle{plain}
\bibliography{inspire}

\begin{thebibliography}{45}
\expandafter\ifx\csname natexlab\endcsname\relax\def\natexlab#1{#1}\fi
\expandafter\ifx\csname bibnamefont\endcsname\relax
  \def\bibnamefont#1{#1}\fi
\expandafter\ifx\csname bibfnamefont\endcsname\relax
  \def\bibfnamefont#1{#1}\fi
\expandafter\ifx\csname citenamefont\endcsname\relax
  \def\citenamefont#1{#1}\fi
\expandafter\ifx\csname url\endcsname\relax
  \def\url#1{\texttt{#1}}\fi
\expandafter\ifx\csname urlprefix\endcsname\relax\def\urlprefix{URL }\fi
\providecommand{\bibinfo}[2]{#2}
\providecommand{\eprint}[2][]{\url{#2}}

\bibitem[{\citenamefont{Skokov et~al.}(2009)\citenamefont{Skokov, Illarionov,
  and Toneev}}]{Skokov:2009qp}
\bibinfo{author}{\bibfnamefont{V.}~\bibnamefont{Skokov}},
  \bibinfo{author}{\bibfnamefont{A.~{\relax Yu}.} \bibnamefont{Illarionov}},
  \bibnamefont{and} \bibinfo{author}{\bibfnamefont{V.}~\bibnamefont{Toneev}},
  \bibinfo{journal}{Int. J. Mod. Phys.} \textbf{\bibinfo{volume}{A24}},
  \bibinfo{pages}{5925} (\bibinfo{year}{2009}), \eprint{0907.1396}.

\bibitem[{\citenamefont{Voronyuk et~al.}(2011)\citenamefont{Voronyuk, Toneev,
  Cassing, Bratkovskaya, Konchakovski, and Voloshin}}]{Voronyuk:2011jd}
\bibinfo{author}{\bibfnamefont{V.}~\bibnamefont{Voronyuk}},
  \bibinfo{author}{\bibfnamefont{V.~D.} \bibnamefont{Toneev}},
  \bibinfo{author}{\bibfnamefont{W.}~\bibnamefont{Cassing}},
  \bibinfo{author}{\bibfnamefont{E.~L.} \bibnamefont{Bratkovskaya}},
  \bibinfo{author}{\bibfnamefont{V.~P.} \bibnamefont{Konchakovski}},
  \bibnamefont{and} \bibinfo{author}{\bibfnamefont{S.~A.}
  \bibnamefont{Voloshin}}, \bibinfo{journal}{Phys. Rev.}
  \textbf{\bibinfo{volume}{C83}}, \bibinfo{pages}{054911}
  (\bibinfo{year}{2011}), \eprint{1103.4239}.

\bibitem[{\citenamefont{Bzdak and Skokov}(2012)}]{Bzdak:2011yy}
\bibinfo{author}{\bibfnamefont{A.}~\bibnamefont{Bzdak}} \bibnamefont{and}
  \bibinfo{author}{\bibfnamefont{V.}~\bibnamefont{Skokov}},
  \bibinfo{journal}{Phys. Lett.} \textbf{\bibinfo{volume}{B710}},
  \bibinfo{pages}{171} (\bibinfo{year}{2012}), \eprint{1111.1949}.

\bibitem[{\citenamefont{Deng and Huang}(2012)}]{Deng:2012pc}
\bibinfo{author}{\bibfnamefont{W.-T.} \bibnamefont{Deng}} \bibnamefont{and}
  \bibinfo{author}{\bibfnamefont{X.-G.} \bibnamefont{Huang}},
  \bibinfo{journal}{Phys. Rev.} \textbf{\bibinfo{volume}{C85}},
  \bibinfo{pages}{044907} (\bibinfo{year}{2012}), \eprint{1201.5108}.

\bibitem[{\citenamefont{Kharzeev et~al.}(2013)\citenamefont{Kharzeev,
  Landsteiner, Schmitt, and Yee}}]{Kharzeev:2012ph}
\bibinfo{author}{\bibfnamefont{D.~E.} \bibnamefont{Kharzeev}},
  \bibinfo{author}{\bibfnamefont{K.}~\bibnamefont{Landsteiner}},
  \bibinfo{author}{\bibfnamefont{A.}~\bibnamefont{Schmitt}}, \bibnamefont{and}
  \bibinfo{author}{\bibfnamefont{H.-U.} \bibnamefont{Yee}},
  \bibinfo{journal}{Lect. Notes Phys.} \textbf{\bibinfo{volume}{871}},
  \bibinfo{pages}{1} (\bibinfo{year}{2013}), \eprint{1211.6245}.

\bibitem[{\citenamefont{Kharzeev et~al.}(2008)\citenamefont{Kharzeev, McLerran,
  and Warringa}}]{Kharzeev:2007jp}
\bibinfo{author}{\bibfnamefont{D.~E.} \bibnamefont{Kharzeev}},
  \bibinfo{author}{\bibfnamefont{L.~D.} \bibnamefont{McLerran}},
  \bibnamefont{and} \bibinfo{author}{\bibfnamefont{H.~J.}
  \bibnamefont{Warringa}}, \bibinfo{journal}{Nucl. Phys.}
  \textbf{\bibinfo{volume}{A803}}, \bibinfo{pages}{227} (\bibinfo{year}{2008}),
  \eprint{0711.0950}.

\bibitem[{\citenamefont{Fukushima et~al.}(2008)\citenamefont{Fukushima,
  Kharzeev, and Warringa}}]{Fukushima:2008xe}
\bibinfo{author}{\bibfnamefont{K.}~\bibnamefont{Fukushima}},
  \bibinfo{author}{\bibfnamefont{D.~E.} \bibnamefont{Kharzeev}},
  \bibnamefont{and} \bibinfo{author}{\bibfnamefont{H.~J.}
  \bibnamefont{Warringa}}, \bibinfo{journal}{Phys. Rev.}
  \textbf{\bibinfo{volume}{D78}}, \bibinfo{pages}{074033}
  (\bibinfo{year}{2008}), \eprint{0808.3382}.

\bibitem[{\citenamefont{Tuchin}(2010)}]{Tuchin:2010vs}
\bibinfo{author}{\bibfnamefont{K.}~\bibnamefont{Tuchin}},
  \bibinfo{journal}{Phys. Rev.} \textbf{\bibinfo{volume}{C82}},
  \bibinfo{pages}{034904} (\bibinfo{year}{2010}), \bibinfo{note}{[Erratum:
  Phys. Rev.C83,039903(2011)]}, \eprint{1006.3051}.

\bibitem[{\citenamefont{Tuchin}(2013{\natexlab{a}})}]{Tuchin:2012mf}
\bibinfo{author}{\bibfnamefont{K.}~\bibnamefont{Tuchin}},
  \bibinfo{journal}{Phys. Rev.} \textbf{\bibinfo{volume}{C87}},
  \bibinfo{pages}{024912} (\bibinfo{year}{2013}{\natexlab{a}}),
  \eprint{1206.0485}.

\bibitem[{\citenamefont{Tuchin}(2013{\natexlab{b}})}]{Tuchin:2013ie}
\bibinfo{author}{\bibfnamefont{K.}~\bibnamefont{Tuchin}},
  \bibinfo{journal}{Adv. High Energy Phys.} \textbf{\bibinfo{volume}{2013}},
  \bibinfo{pages}{490495} (\bibinfo{year}{2013}{\natexlab{b}}),
  \eprint{1301.0099}.

\bibitem[{\citenamefont{Mohapatra et~al.}(2011)\citenamefont{Mohapatra, Saumia,
  and Srivastava}}]{Mohapatra:2011ku}
\bibinfo{author}{\bibfnamefont{R.~K.} \bibnamefont{Mohapatra}},
  \bibinfo{author}{\bibfnamefont{P.~S.} \bibnamefont{Saumia}},
  \bibnamefont{and} \bibinfo{author}{\bibfnamefont{A.~M.}
  \bibnamefont{Srivastava}}, \bibinfo{journal}{Mod. Phys. Lett.}
  \textbf{\bibinfo{volume}{A26}}, \bibinfo{pages}{2477} (\bibinfo{year}{2011}),
  \eprint{1102.3819}.

\bibitem[{\citenamefont{Bali et~al.}(2014{\natexlab{a}})\citenamefont{Bali,
  Bruckmann, Endr\H{o}di, and Sch{\"a}fer}}]{Bali:2013owa}
\bibinfo{author}{\bibfnamefont{G.~S.} \bibnamefont{Bali}},
  \bibinfo{author}{\bibfnamefont{F.}~\bibnamefont{Bruckmann}},
  \bibinfo{author}{\bibfnamefont{G.}~\bibnamefont{Endr\H{o}di}},
  \bibnamefont{and}
  \bibinfo{author}{\bibfnamefont{A.}~\bibnamefont{Sch{\"a}fer}},
  \bibinfo{journal}{Phys. Rev. Lett.} \textbf{\bibinfo{volume}{112}},
  \bibinfo{pages}{042301} (\bibinfo{year}{2014}{\natexlab{a}}),
  \eprint{1311.2559}.

\bibitem[{\citenamefont{Elmfors et~al.}(1993)\citenamefont{Elmfors, Persson,
  and Skagerstam}}]{Elmfors:1993wj}
\bibinfo{author}{\bibfnamefont{P.}~\bibnamefont{Elmfors}},
  \bibinfo{author}{\bibfnamefont{D.}~\bibnamefont{Persson}}, \bibnamefont{and}
  \bibinfo{author}{\bibfnamefont{B.-S.} \bibnamefont{Skagerstam}},
  \bibinfo{journal}{Phys. Rev. Lett.} \textbf{\bibinfo{volume}{71}},
  \bibinfo{pages}{480} (\bibinfo{year}{1993}), \eprint{hep-th/9305004}.

\bibitem[{\citenamefont{Endr\H{o}di}(2013)}]{Endrodi:2013cs}
\bibinfo{author}{\bibfnamefont{G.}~\bibnamefont{Endr\H{o}di}},
  \bibinfo{journal}{JHEP} \textbf{\bibinfo{volume}{04}}, \bibinfo{pages}{023}
  (\bibinfo{year}{2013}), \eprint{1301.1307}.

\bibitem[{\citenamefont{Bali et~al.}(2013)\citenamefont{Bali, Bruckmann,
  Endr\H{o}di, Gruber, and Sch{\"a}fer}}]{Bali:2013esa}
\bibinfo{author}{\bibfnamefont{G.~S.} \bibnamefont{Bali}},
  \bibinfo{author}{\bibfnamefont{F.}~\bibnamefont{Bruckmann}},
  \bibinfo{author}{\bibfnamefont{G.}~\bibnamefont{Endr\H{o}di}},
  \bibinfo{author}{\bibfnamefont{F.}~\bibnamefont{Gruber}}, \bibnamefont{and}
  \bibinfo{author}{\bibfnamefont{A.}~\bibnamefont{Sch{\"a}fer}},
  \bibinfo{journal}{JHEP} \textbf{\bibinfo{volume}{04}}, \bibinfo{pages}{130}
  (\bibinfo{year}{2013}), \eprint{1303.1328}.

\bibitem[{\citenamefont{Bali et~al.}(2014{\natexlab{b}})\citenamefont{Bali,
  Bruckmann, Endr\H{o}di, and Sch{\"a}fer}}]{Bali:2013txa}
\bibinfo{author}{\bibfnamefont{G.~S.} \bibnamefont{Bali}},
  \bibinfo{author}{\bibfnamefont{F.}~\bibnamefont{Bruckmann}},
  \bibinfo{author}{\bibfnamefont{G.}~\bibnamefont{Endr\H{o}di}},
  \bibnamefont{and}
  \bibinfo{author}{\bibfnamefont{A.}~\bibnamefont{Sch{\"a}fer}},
  \bibinfo{journal}{PoS} \textbf{\bibinfo{volume}{LATTICE2013}},
  \bibinfo{pages}{182} (\bibinfo{year}{2014}{\natexlab{b}}),
  \eprint{1310.8145}.

\bibitem[{\citenamefont{Bonati et~al.}(2013)\citenamefont{Bonati, D'Elia,
  Mariti, Negro, and Sanfilippo}}]{Bonati:2013lca}
\bibinfo{author}{\bibfnamefont{C.}~\bibnamefont{Bonati}},
  \bibinfo{author}{\bibfnamefont{M.}~\bibnamefont{D'Elia}},
  \bibinfo{author}{\bibfnamefont{M.}~\bibnamefont{Mariti}},
  \bibinfo{author}{\bibfnamefont{F.}~\bibnamefont{Negro}}, \bibnamefont{and}
  \bibinfo{author}{\bibfnamefont{F.}~\bibnamefont{Sanfilippo}},
  \bibinfo{journal}{Phys. Rev. Lett.} \textbf{\bibinfo{volume}{111}},
  \bibinfo{pages}{182001} (\bibinfo{year}{2013}), \eprint{1307.8063}.

\bibitem[{\citenamefont{Bonati et~al.}(2014)\citenamefont{Bonati, D'Elia,
  Mariti, Negro, and Sanfilippo}}]{Bonati:2013vba}
\bibinfo{author}{\bibfnamefont{C.}~\bibnamefont{Bonati}},
  \bibinfo{author}{\bibfnamefont{M.}~\bibnamefont{D'Elia}},
  \bibinfo{author}{\bibfnamefont{M.}~\bibnamefont{Mariti}},
  \bibinfo{author}{\bibfnamefont{F.}~\bibnamefont{Negro}}, \bibnamefont{and}
  \bibinfo{author}{\bibfnamefont{F.}~\bibnamefont{Sanfilippo}},
  \bibinfo{journal}{Phys. Rev.} \textbf{\bibinfo{volume}{D89}},
  \bibinfo{pages}{054506} (\bibinfo{year}{2014}), \eprint{1310.8656}.

\bibitem[{\citenamefont{Levkova and DeTar}(2014)}]{Levkova:2013qda}
\bibinfo{author}{\bibfnamefont{L.}~\bibnamefont{Levkova}} \bibnamefont{and}
  \bibinfo{author}{\bibfnamefont{C.}~\bibnamefont{DeTar}},
  \bibinfo{journal}{Phys. Rev. Lett.} \textbf{\bibinfo{volume}{112}},
  \bibinfo{pages}{012002} (\bibinfo{year}{2014}), \eprint{1309.1142}.

\bibitem[{\citenamefont{Bali et~al.}(2014{\natexlab{c}})\citenamefont{Bali,
  Bruckmann, Endr\H{o}di, Katz, and Sch{\"a}fer}}]{Bali:2014kia}
\bibinfo{author}{\bibfnamefont{G.~S.} \bibnamefont{Bali}},
  \bibinfo{author}{\bibfnamefont{F.}~\bibnamefont{Bruckmann}},
  \bibinfo{author}{\bibfnamefont{G.}~\bibnamefont{Endr\H{o}di}},
  \bibinfo{author}{\bibfnamefont{S.~D.} \bibnamefont{Katz}}, \bibnamefont{and}
  \bibinfo{author}{\bibfnamefont{A.}~\bibnamefont{Sch{\"a}fer}},
  \bibinfo{journal}{JHEP} \textbf{\bibinfo{volume}{08}}, \bibinfo{pages}{177}
  (\bibinfo{year}{2014}{\natexlab{c}}), \eprint{1406.0269}.

\bibitem[{\citenamefont{Landau and Lifshitz}(1995)}]{landau:1995em}
\bibinfo{author}{\bibfnamefont{L.}~\bibnamefont{Landau}} \bibnamefont{and}
  \bibinfo{author}{\bibfnamefont{E.}~\bibnamefont{Lifshitz}},
  \bibinfo{journal}{Electrodynamics of Continuous Media}
  (\bibinfo{year}{1995}).

\bibitem[{\citenamefont{Lyutikov and Hadden}(2012)}]{Lyutikov:2011vc}
\bibinfo{author}{\bibfnamefont{M.}~\bibnamefont{Lyutikov}} \bibnamefont{and}
  \bibinfo{author}{\bibfnamefont{S.}~\bibnamefont{Hadden}},
  \bibinfo{journal}{Phys. Rev.} \textbf{\bibinfo{volume}{E85}},
  \bibinfo{pages}{026401} (\bibinfo{year}{2012}), \eprint{1112.0249}.

\bibitem[{\citenamefont{Roy et~al.}(2015)\citenamefont{Roy, Pu, Rezzolla, and
  Rischke}}]{Roy:2015kma}
\bibinfo{author}{\bibfnamefont{V.}~\bibnamefont{Roy}},
  \bibinfo{author}{\bibfnamefont{S.}~\bibnamefont{Pu}},
  \bibinfo{author}{\bibfnamefont{L.}~\bibnamefont{Rezzolla}}, \bibnamefont{and}
  \bibinfo{author}{\bibfnamefont{D.}~\bibnamefont{Rischke}},
  \bibinfo{journal}{Phys. Lett.} \textbf{\bibinfo{volume}{B750}},
  \bibinfo{pages}{45} (\bibinfo{year}{2015}), \eprint{1506.06620}.

\bibitem[{\citenamefont{Roy and Pu}(2015)}]{Roy:2015coa}
\bibinfo{author}{\bibfnamefont{V.}~\bibnamefont{Roy}} \bibnamefont{and}
  \bibinfo{author}{\bibfnamefont{S.}~\bibnamefont{Pu}}, \bibinfo{journal}{Phys.
  Rev.} \textbf{\bibinfo{volume}{C92}}, \bibinfo{pages}{064902}
  (\bibinfo{year}{2015}), \eprint{1508.03761}.

\bibitem[{\citenamefont{Pu et~al.}(2016)\citenamefont{Pu, Roy, Rezzolla, and
  Rischke}}]{Pu:2016ayh}
\bibinfo{author}{\bibfnamefont{S.}~\bibnamefont{Pu}},
  \bibinfo{author}{\bibfnamefont{V.}~\bibnamefont{Roy}},
  \bibinfo{author}{\bibfnamefont{L.}~\bibnamefont{Rezzolla}}, \bibnamefont{and}
  \bibinfo{author}{\bibfnamefont{D.~H.} \bibnamefont{Rischke}}
  (\bibinfo{year}{2016}), \eprint{1602.04953}.

\bibitem[{\citenamefont{Pu and Yang}(2016)}]{Pu:2016bxy}
\bibinfo{author}{\bibfnamefont{S.}~\bibnamefont{Pu}} \bibnamefont{and}
  \bibinfo{author}{\bibfnamefont{D.-L.} \bibnamefont{Yang}}
  (\bibinfo{year}{2016}), \eprint{1602.04954}.

\bibitem[{\citenamefont{Pang et~al.}(2012)\citenamefont{Pang, Wang, and
  Wang}}]{Pang:2012he}
\bibinfo{author}{\bibfnamefont{L.}~\bibnamefont{Pang}},
  \bibinfo{author}{\bibfnamefont{Q.}~\bibnamefont{Wang}}, \bibnamefont{and}
  \bibinfo{author}{\bibfnamefont{X.-N.} \bibnamefont{Wang}},
  \bibinfo{journal}{Phys. Rev.} \textbf{\bibinfo{volume}{C86}},
  \bibinfo{pages}{024911} (\bibinfo{year}{2012}), \eprint{1205.5019}.

\bibitem[{\citenamefont{Pang et~al.}(2015)\citenamefont{Pang, Hatta, Wang, and
  Xiao}}]{Pang:2014ipa}
\bibinfo{author}{\bibfnamefont{L.-G.} \bibnamefont{Pang}},
  \bibinfo{author}{\bibfnamefont{Y.}~\bibnamefont{Hatta}},
  \bibinfo{author}{\bibfnamefont{X.-N.} \bibnamefont{Wang}}, \bibnamefont{and}
  \bibinfo{author}{\bibfnamefont{B.-W.} \bibnamefont{Xiao}},
  \bibinfo{journal}{Phys. Rev.} \textbf{\bibinfo{volume}{D91}},
  \bibinfo{pages}{074027} (\bibinfo{year}{2015}), \eprint{1411.7767}.

\bibitem[{\citenamefont{Felderhof and Kroh}(1999)}]{Felderhof:1999Mhd}
\bibinfo{author}{\bibfnamefont{B.~U.} \bibnamefont{Felderhof}}
  \bibnamefont{and} \bibinfo{author}{\bibfnamefont{H.~J.} \bibnamefont{Kroh}},
  \bibinfo{journal}{The Journal of Chemical Physics}
  \textbf{\bibinfo{volume}{110}} (\bibinfo{year}{1999}).

\bibitem[{\citenamefont{Bors\'anyi et~al.}(2014)\citenamefont{Bors\'anyi,
  Fodor, Hoelbling, Katz, Krieg, and Szab\'o}}]{Borsanyi:2013bia}
\bibinfo{author}{\bibfnamefont{S.}~\bibnamefont{Bors\'anyi}},
  \bibinfo{author}{\bibfnamefont{Z.}~\bibnamefont{Fodor}},
  \bibinfo{author}{\bibfnamefont{C.}~\bibnamefont{Hoelbling}},
  \bibinfo{author}{\bibfnamefont{S.~D.} \bibnamefont{Katz}},
  \bibinfo{author}{\bibfnamefont{S.}~\bibnamefont{Krieg}}, \bibnamefont{and}
  \bibinfo{author}{\bibfnamefont{K.~K.} \bibnamefont{Szab\'o}},
  \bibinfo{journal}{Phys. Lett.} \textbf{\bibinfo{volume}{B730}},
  \bibinfo{pages}{99} (\bibinfo{year}{2014}), \eprint{1309.5258}.

\bibitem[{\citenamefont{McLerran and Skokov}(2014)}]{McLerran:2013hla}
\bibinfo{author}{\bibfnamefont{L.}~\bibnamefont{McLerran}} \bibnamefont{and}
  \bibinfo{author}{\bibfnamefont{V.}~\bibnamefont{Skokov}},
  \bibinfo{journal}{Nucl. Phys.} \textbf{\bibinfo{volume}{A929}},
  \bibinfo{pages}{184} (\bibinfo{year}{2014}), \eprint{1305.0774}.

\bibitem[{\citenamefont{Li et~al.}(2016)\citenamefont{Li, Sheng, and
  Wang}}]{Li:2016tel}
\bibinfo{author}{\bibfnamefont{H.}~\bibnamefont{Li}},
  \bibinfo{author}{\bibfnamefont{X.-l.} \bibnamefont{Sheng}}, \bibnamefont{and}
  \bibinfo{author}{\bibfnamefont{Q.}~\bibnamefont{Wang}}
  (\bibinfo{year}{2016}), \eprint{1602.02223}.

\bibitem[{\citenamefont{Schwinger}(1951)}]{Schwinger:1951nm}
\bibinfo{author}{\bibfnamefont{J.~S.} \bibnamefont{Schwinger}},
  \bibinfo{journal}{Phys. Rev.} \textbf{\bibinfo{volume}{82}},
  \bibinfo{pages}{664} (\bibinfo{year}{1951}).

\bibitem[{\citenamefont{Tuchin}(2013{\natexlab{c}})}]{Tuchin:2013apa}
\bibinfo{author}{\bibfnamefont{K.}~\bibnamefont{Tuchin}},
  \bibinfo{journal}{Phys. Rev.} \textbf{\bibinfo{volume}{C88}},
  \bibinfo{pages}{024911} (\bibinfo{year}{2013}{\natexlab{c}}),
  \eprint{1305.5806}.

\bibitem[{\citenamefont{Zhong et~al.}(2015)\citenamefont{Zhong, Yang, Cai, and
  Feng}}]{Zhong:2014sua}
\bibinfo{author}{\bibfnamefont{Y.}~\bibnamefont{Zhong}},
  \bibinfo{author}{\bibfnamefont{C.-B.} \bibnamefont{Yang}},
  \bibinfo{author}{\bibfnamefont{X.}~\bibnamefont{Cai}}, \bibnamefont{and}
  \bibinfo{author}{\bibfnamefont{S.-Q.} \bibnamefont{Feng}},
  \bibinfo{journal}{Chin. Phys.} \textbf{\bibinfo{volume}{C39}},
  \bibinfo{pages}{104105} (\bibinfo{year}{2015}), \eprint{1410.6349}.

\bibitem[{\citenamefont{Gursoy et~al.}(2014)\citenamefont{Gursoy, Kharzeev, and
  Rajagopal}}]{Gursoy:2014aka}
\bibinfo{author}{\bibfnamefont{U.}~\bibnamefont{Gursoy}},
  \bibinfo{author}{\bibfnamefont{D.}~\bibnamefont{Kharzeev}}, \bibnamefont{and}
  \bibinfo{author}{\bibfnamefont{K.}~\bibnamefont{Rajagopal}},
  \bibinfo{journal}{Phys. Rev.} \textbf{\bibinfo{volume}{C89}},
  \bibinfo{pages}{054905} (\bibinfo{year}{2014}), \eprint{1401.3805}.

\bibitem[{\citenamefont{Zakharov}(2014)}]{Zakharov:2014dia}
\bibinfo{author}{\bibfnamefont{B.~G.} \bibnamefont{Zakharov}},
  \bibinfo{journal}{Phys. Lett.} \textbf{\bibinfo{volume}{B737}},
  \bibinfo{pages}{262} (\bibinfo{year}{2014}), \eprint{1404.5047}.

\bibitem[{\citenamefont{Gupta}(2004)}]{Gupta:2003zh}
\bibinfo{author}{\bibfnamefont{S.}~\bibnamefont{Gupta}},
  \bibinfo{journal}{Phys. Lett.} \textbf{\bibinfo{volume}{B597}},
  \bibinfo{pages}{57} (\bibinfo{year}{2004}), \eprint{hep-lat/0301006}.

\bibitem[{\citenamefont{Qin}(2015)}]{Qin:2013aaa}
\bibinfo{author}{\bibfnamefont{S.-x.} \bibnamefont{Qin}},
  \bibinfo{journal}{Phys. Lett.} \textbf{\bibinfo{volume}{B742}},
  \bibinfo{pages}{358} (\bibinfo{year}{2015}), \eprint{1307.4587}.

\bibitem[{\citenamefont{Greif et~al.}(2014)\citenamefont{Greif, Bouras,
  Greiner, and Xu}}]{Greif:2014oia}
\bibinfo{author}{\bibfnamefont{M.}~\bibnamefont{Greif}},
  \bibinfo{author}{\bibfnamefont{I.}~\bibnamefont{Bouras}},
  \bibinfo{author}{\bibfnamefont{C.}~\bibnamefont{Greiner}}, \bibnamefont{and}
  \bibinfo{author}{\bibfnamefont{Z.}~\bibnamefont{Xu}}, \bibinfo{journal}{Phys.
  Rev.} \textbf{\bibinfo{volume}{D90}}, \bibinfo{pages}{094014}
  (\bibinfo{year}{2014}), \eprint{1408.7049}.

\bibitem[{\citenamefont{Schenke et~al.}(2010)\citenamefont{Schenke, Jeon, and
  Gale}}]{Schenke:2010nt}
\bibinfo{author}{\bibfnamefont{B.}~\bibnamefont{Schenke}},
  \bibinfo{author}{\bibfnamefont{S.}~\bibnamefont{Jeon}}, \bibnamefont{and}
  \bibinfo{author}{\bibfnamefont{C.}~\bibnamefont{Gale}},
  \bibinfo{journal}{Phys. Rev.} \textbf{\bibinfo{volume}{C82}},
  \bibinfo{pages}{014903} (\bibinfo{year}{2010}), \eprint{1004.1408}.

\bibitem[{\citenamefont{Tuchin}(2012)}]{Tuchin:2011jw}
\bibinfo{author}{\bibfnamefont{K.}~\bibnamefont{Tuchin}}, \bibinfo{journal}{J.
  Phys.} \textbf{\bibinfo{volume}{G39}}, \bibinfo{pages}{025010}
  (\bibinfo{year}{2012}), \eprint{1108.4394}.

\bibitem[{\citenamefont{Critelli et~al.}(2014)\citenamefont{Critelli, Finazzo,
  Zaniboni, and Noronha}}]{Critelli:2014kra}
\bibinfo{author}{\bibfnamefont{R.}~\bibnamefont{Critelli}},
  \bibinfo{author}{\bibfnamefont{S.~I.} \bibnamefont{Finazzo}},
  \bibinfo{author}{\bibfnamefont{M.}~\bibnamefont{Zaniboni}}, \bibnamefont{and}
  \bibinfo{author}{\bibfnamefont{J.}~\bibnamefont{Noronha}},
  \bibinfo{journal}{Phys. Rev.} \textbf{\bibinfo{volume}{D90}},
  \bibinfo{pages}{066006} (\bibinfo{year}{2014}), \eprint{1406.6019}.

\bibitem[{\citenamefont{Romatschke and Romatschke}(2007)}]{Romatschke:2007mq}
\bibinfo{author}{\bibfnamefont{P.}~\bibnamefont{Romatschke}} \bibnamefont{and}
  \bibinfo{author}{\bibfnamefont{U.}~\bibnamefont{Romatschke}},
  \bibinfo{journal}{Phys. Rev. Lett.} \textbf{\bibinfo{volume}{99}},
  \bibinfo{pages}{172301} (\bibinfo{year}{2007}), \eprint{0706.1522}.

\bibitem[{\citenamefont{Song and Heinz}(2008)}]{Song:2007fn}
\bibinfo{author}{\bibfnamefont{H.}~\bibnamefont{Song}} \bibnamefont{and}
  \bibinfo{author}{\bibfnamefont{U.~W.} \bibnamefont{Heinz}},
  \bibinfo{journal}{Phys. Lett.} \textbf{\bibinfo{volume}{B658}},
  \bibinfo{pages}{279} (\bibinfo{year}{2008}), \eprint{0709.0742}.

\end{thebibliography}

\end{acknowledgments}
\end{document}